\documentclass[usenatbib]{mnras}

\usepackage{natbib}
\usepackage{delarray}
\usepackage{xcolor}
\usepackage{amsmath}
\usepackage{amssymb}
\usepackage{here}
\usepackage{graphicx}
\usepackage[utf8]{inputenc}
\usepackage{float}
\usepackage{epsfig}
\usepackage{tensor}
\usepackage{xspace}
\usepackage{multirow}

\usepackage{newtxtext,newtxmath}

\usepackage[T1]{fontenc}
\usepackage{ae,aecompl}

\newcommand{\appropto}{\mathrel{\vcenter{
  \offinterlineskip\halign{\hfil$##$\cr
    \propto\cr\noalign{\kern2pt}\sim\cr\noalign{\kern-2pt}}}}}

\newcommand{\ord}{\mathcal{O}}
\newcommand{\mach}{\mathcal{M}}

\newcommand{\be}{\begin{equation}} \newcommand{\ee}{\end{equation}}

\newcommand{\solarmass}{\mathrm{M}_{\rm \sun}}
\newcommand{\msun}{\solarmass}

\newcommand{\dderiv}{\mathrm{d}}

\newcommand{\red}[1]{\textcolor{red}{#1}}

\newcommand{\green}[1]{\textcolor{green}{#1}}

\newcommand{\orange}[1]{\textcolor{orange}{#1}}

\newcommand{\acknowledgments}{\begin{small}\section*{Acknowledgments}\end{small}}
\newcommand\altaffilmark[1]{$^{#1}$}
\newcommand\altaffiltext[1]{$^{#1}$}

\newcommand{\myquote}[1]{``#1''}

\usepackage[normalem]{ulem}

\title[Extragalactic IMF variations vs universality]{Is it possible to reconcile extragalactic IMF variations with a universal Milky Way IMF?}

\author[Guszejnov, Hopkins \&\ Graus]{
\parbox[t]{\textwidth}{ D\'avid Guszejnov\altaffilmark{1}\thanks{E-mail:guszejnov.david@gmail.com}, Philip F. Hopkins\altaffilmark{2} and Andrew S. Graus\altaffilmark{1}}
\vspace*{6pt} \\
\altaffiltext{1}{Department of Astronomy, The University of Texas at Austin, Austin, TX 78712, USA} \\
\altaffiltext{2}{TAPIR, MC 350-17, California Institute of Technology, Pasadena, CA 91125, USA} \\
}

\date{To be submitted to MNRAS, \today \vspace{-0.6cm}}
\begin{document}
\maketitle
\label{firstpage}

\begin{abstract}
One of the most robust observations of the stellar initial mass function (IMF) is its near-universality in the Milky Way and neighboring galaxies. But recent observations of early-type galaxies can be interpreted to imply a \myquote{bottom-heavy} IMF, while others of ultra-faint dwarfs could imply a \myquote{top-heavy} IMF. This would impose powerful constraints on star formation models. We explore what sort of \myquote{cloud-scale} IMF models could possibly satisfy these constraints. We utilize simulated galaxies which reproduce (broadly) the observed galaxy properties, while they also provide the detailed star formation history and properties of each progenitor star-forming cloud. We then consider generic models where the characteristic mass of the IMF is some {\em arbitrary} power-law function of progenitor cloud properties, along with well-known literature IMF models which scale with Jeans mass, \myquote{turbulent Bonnor-Ebert mass}, temperature, the opacity limit, metallicity, or the \myquote{protostellar heating mass}. We show that no IMF models currently in the literature -- nor {\em any} model where the turnover mass is an arbitrary power-law function of a combination of cloud temperature/density/size/metallicity/velocity dispersion/magnetic field -- can reproduce the claimed IMF variation in ellipticals or dwarfs without severely violating observational constraints in the Milky Way. Specifically, they predict too much variation in the \myquote{extreme} environments of the Galaxy, compared to that observed. Either the IMF varies in a more complicated manner, or alternative interpretations of the extragalactic observations must be explored.
\end{abstract}

\begin{keywords}
stars: formation -- turbulence -- galaxies: star formation -- cosmology: theory
\vspace{-1.0cm}
\end{keywords}

\section{Introduction}\label{sec:intro}

The mass distribution of stars at formation (often called the initial mass function or \textit{IMF}) is a key part of cosmic evolution as it affects essentially all astrophysical scales. A key finding regarding the IMF is its apparent universality in the Milky Way and its satellite galaxies \citep[see the reviews of][]{Chabrier_review_2003,imf_review,imf_universality,SF_big_problems}, regardless of the locations and age of the observed population with a few possible outliers \citep[e.g.][]{Espinoza_2009_Arches_IMF}. While the IMF appears to be universal in the Galaxy, recent observations of the extragalactic IMF have been extrapolated to imply significant variations. Recent studies have looked at the centers of massive early type galaxies (ETGs) and have seen an apparent excess of low mass stars, a \myquote{bottom-heavy IMF} \citep[see][]{Conroy_vanDokkum_ellipticals,vanDokkum_conroy_2011,ConryVanDokkum_IMFvar_2012,Treu_galaxy_IMF_2010,Sonnenfeld_Treu_2015_IMF,Cappellari_IMF_var_2012,Posacki_Cappellari_2015_IMF,navarro_2015_imf_relic,navarro_2015_imf_relation}. One must be cautious as these measurements use fundamentally different methods (e.g. Stellar Population Synthesis, SPS) than those used for IMF measurement in the Milky Way (e.g. star counting). For more details see the recent review of \citealt{Hopkins_A_2018_IMF_obs_review}. This means that various interpretation of these extragalactic results can lead to different implied IMFs. Furthermore, several other studies conflict with the claimed variations \citep[e.g.][]{Smith_2014_dynamical_IMF_novar,Smith_2015_dynamical_IMF_novar,Smith_2017_dynamical_IMF_novar,Collier_ETG_no_IMF_var_2018}. Meanwhile observations relying on star counts in Ultra Faint Dwarf Galaxies imply an overabundance of high mass stars, a \myquote{top-heavy IMF} \citep[see][]{Geha_2013_UFD_IMF,Gennaro_2018_UFD_IMF1}. Note, that due to large uncertainties in the results, MW-like IMFs are not entirely ruled out by many of these observations \citep[e.g.][]{Offner_2016_IMF_progress}, and some UFD galaxies have IMFs consistent with a MW IMF, which means that the previously observed variations could be due to observational artefacts \citep{ElBadry_2017_dwarfIMF_universal,Gennaro_2018_UFD_IMF2}.

Nonetheless, several theoretical models have been proposed to explain the claimed IMF variations. In fact, analytic sonic mass/turbulent Bonnor-Ebert mass models\footnote{In these models the IMF is regulated by isothermal turbulence, the higher the turbulent velocity dispersion the more the clouds fragment, which leads to more low mass stars.} \citep[e.g.][]{hc08,hc_2013, core_imf} provide a remarkably good fit to the observed bottom-heavy IMF in ETGs. Several empirical models have also been proposed including ones that tie the IMF variations to metallicity \citep{navarro_2015_imf_relation}\footnote{Note that the observed IMF variations in ETGs correlate not only with metallicity but with the galactic scale velocity dispersion as well, see \cite{Zhou_2018_SDSS_MANGA_IMF_var}.}. However, these models all have trouble explaining the apparent universality of the IMF in the Galaxy. For example, \cite{guszejnov_imf_var} have shown that the above mentioned sonic-mass based models predict too much IMF variation {\em within the MW} -- dense, high Mach number regions like the Arches cluster, the \myquote{Brick} cloud and the galactic nucleus should have bottom heavy IMFs, while observations indicate a slightly top-heavy IMF \citep{Espinoza_2009_Arches_IMF,Hussmann_2012_Quintuplet_IMF}. Furthermore, IMF measurements using star counting have consistently found no sign of metallicity dependence in globular clusters from super solar metallicities down to $[Z/H]\sim -2$ (e.g., \citealt{De_Marchi_2010_globular_clusters}, see \citealt{imf_review} and references therein). In fact, the most successful IMF models in the MW rely on heating via protostellar radiation to self-regulate the IMF with a turnover mass that depend {\em exceptionally weakly} on the environmental properties \citep[see][]{krumholz_stellar_mass_origin}. A significant problem with these models is that - due to their weak dependence on gas properties - they are likely unable to reproduce the claimed IMF variations even in extreme environments.

In this paper, we explore whether it is possible to reconcile the claimed IMF variations in ETGs and UFDs with the apparent IMF universality in the Milky Way. Similar to \cite{guszejnov_imf_var} we investigate IMF models that assume that the IMF of a stellar population is set by the properties of the progenitor cloud out of which they form. We gather cloud properties from high resolution cosmological simulations of different types of galaxies, allowing us to predict the properties of the \emph{entire stellar population} of the galaxy for any IMF model. 

The paper is organized as follows. In \S~\ref{sec:IMF_char_mass} we introduce a simple 1-parameter analytic model (based on \citealt{Maschberger_2013_IMF_parametrization}) to map the observed \myquote{IMF slopes} to actual mass functions. In \S~\ref{sec:constraints} we quantify both the observed IMF universality in the MW and the observed variations in UFDs and ETGs within the framework of our 1-parameter model. \S~\ref{sec:sim} details the properties of the simulated galaxies we use for this study, while \S~\ref{sec:imf_models} details the specific IMF models we investigate. We present our final results in \S~\ref{sec:results}. 






\section{Model and Methods}\label{sec:methods}

\subsection{IMF slope and characteristic mass}\label{sec:IMF_char_mass}

Observations relying on the integrated spectra of galaxies (SPS modeling) are currently unable to probe the IMF in its entire mass range. Instead they constrain the relative number for a few select types of stars, effectively calculating the \myquote{slope} of the IMF in relatively small mass ranges. As different measurements probe slightly different regions of the IMF \citep{Hopkins_A_2018_IMF_obs_review}, it is necessary to find a model that allows one to compare between these measurements.

In this paper we use a simplified version of the $L_3$ parametric IMF model of \cite{Maschberger_2013_IMF_parametrization}. In the $L_3$ model the IMF has the following form:
\be 
\frac{\dderiv N}{\dderiv M}=L_3\left(M,\alpha,\beta\right)\equiv A\left(\frac{M}{\mu}\right)^{-\alpha}\left(1+\left(\frac{M}{\mu}\right)^{1-\alpha}\right)^{-\beta}
\label{eq:L3_IMF}
\ee
In the low and high mass limits this simplifies to power laws with $-\alpha$ and $-\alpha-\beta\left(1-\alpha\right)$ slopes respectively. The characteristic mass scale is $\mu$, this is where the function transitions between the two limits. Note that $A$ is just a normalization constant that depends on $\alpha$, $\beta$ and $\mu$, as well as $m_l$ and $m_u$, the low and high mass cut-offs of the IMF for which we use $m_l=0.01\,\msun$ and $m_u=150\,\msun$.

\begin{figure}
\begin {center}
\includegraphics[width=\linewidth]{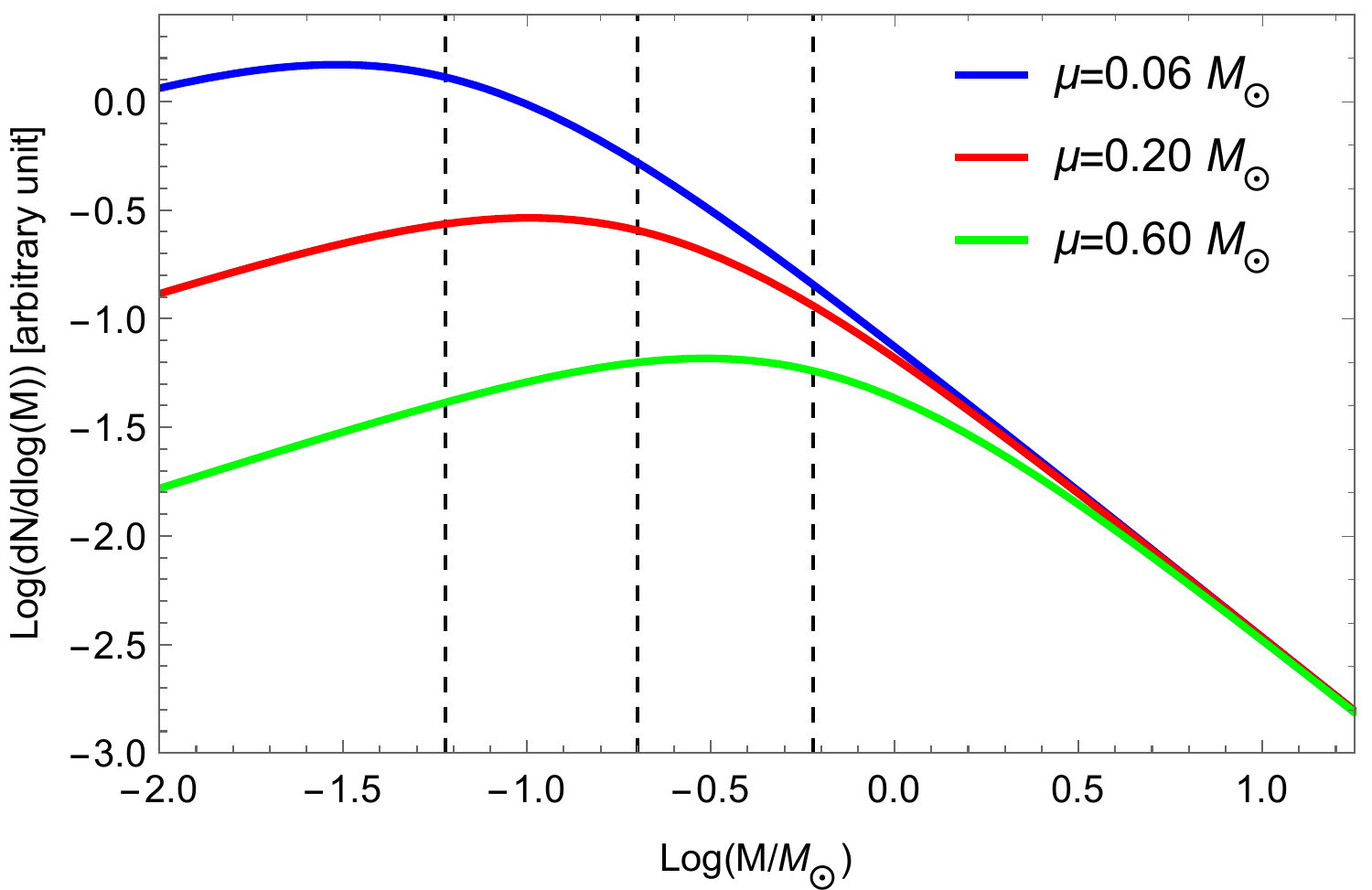}
\vspace{-0.5cm}
\caption{Effects of different characteristic masses $\mu$ in the one-parameter IMF model we adopt in this paper. The characteristic scale $\mu$ essentially sets the mass scale where the PDF deviates from the high-mass power-law behavior.}
\label{fig:imf_var}
\vspace{-0.5cm}
\end {center}
\end{figure}

Since most observations only measure a single slope of the IMF, it is necessary to reduce the number of parameters for our IMF model. In this paper we adopt $\alpha=2.3$ and $\beta=1.4$, which are the canonical fit values for the MW IMF. The adoption of these \myquote{fixed} slopes is further motivated by the fact that most scale-free structure formation processes naturally produce a $-2$ slope in the mass function \citep{guszejnov_scaling_laws}. These parameters lead to the one parameter IMF model that we adopt for the rest of the paper, where
\be 
\frac{\dderiv N}{\dderiv M}\propto\left(\frac{M}{\mu}\right)^{-2.3}\left(1+\left(\frac{M}{\mu}\right)^{-1.3}\right)^{-1.4}
\label{eq:1param_IMF}
\ee
This leaves the characteristic mass scale $\mu$ as the only free parameter, so our model essentially assumes that the IMF has a universal shape that can only be shifted to lower or higher masses (see Figure \ref{fig:imf_var}). As observations only constrain the IMF slope in a small dynamic range, such a one-parameter IMF can fit the observations.

\begin{figure}
\begin {center}
\includegraphics[width=\linewidth]{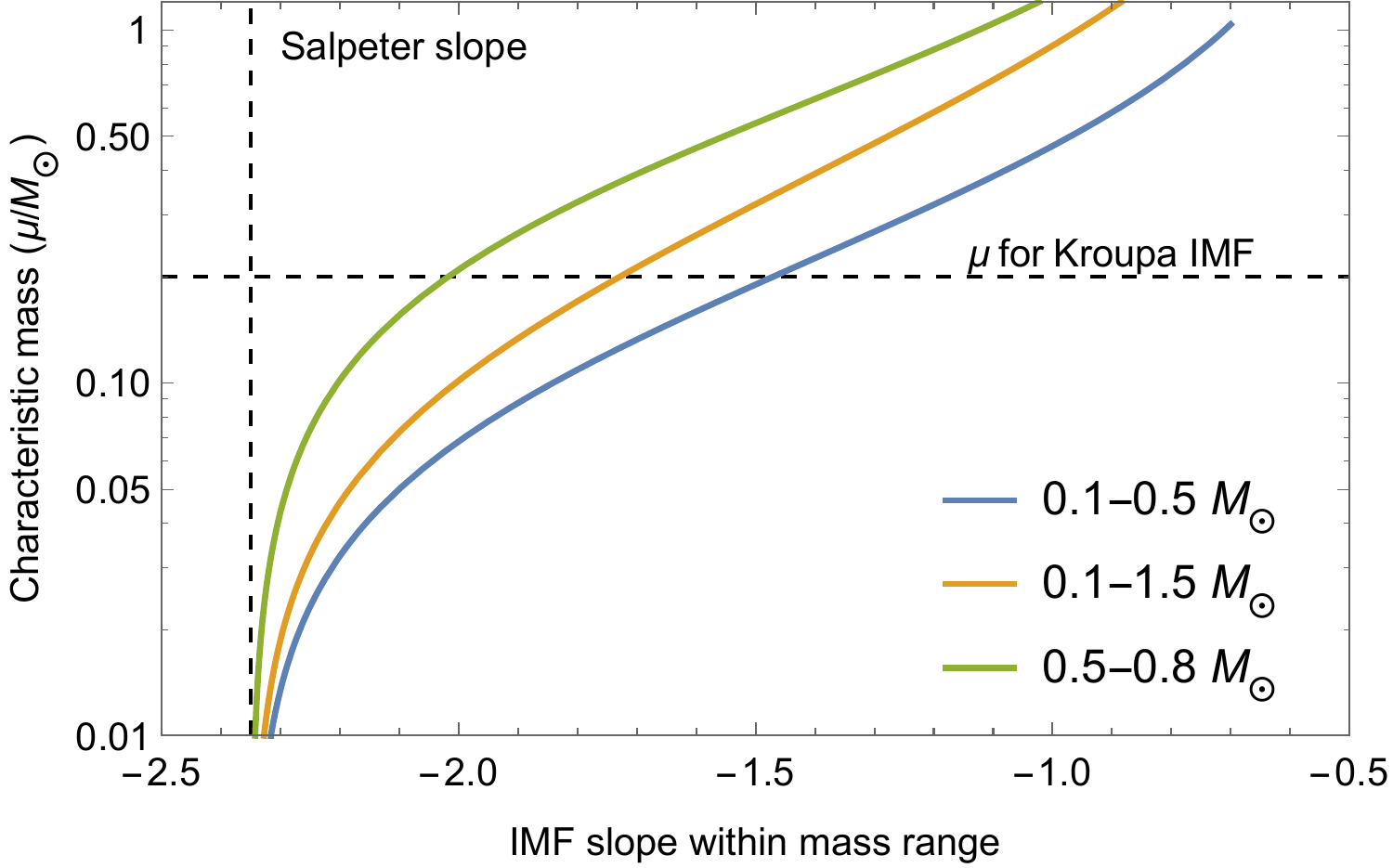}
\vspace{-0.5cm}
\caption{IMF slope and characteristic mass, based on where the slope is calculated using the one-parameter IMF model of Eq. \ref{eq:1param_IMF}. Since most observations of the extragalactic IMF measure the \myquote{slope} in different mass ranges, it is necessary to calculate the appropriate characteristic mass $\mu$ to interpret them (see Figure \ref{fig:imf_var} for the effects of $\mu$ on the IMF).}
\label{fig:mu_vs_slope}
\vspace{-0.5cm}
\end {center}
\end{figure}

Armed with this simple, one-parameter model we can create a one-to-one mapping between the slopes measured in different mass ranges and the characteristic mass (the \myquote{peak}/\myquote{turnover mass}) of the IMF. Figure \ref{fig:mu_vs_slope} shows that the inferred characteristic mass is sensitive to the probed mass range, so one should be cautious when trying to compare different observations.

One of the aims of this paper is to explore the space of possible IMF models and identify what regions of this space would satisfy observational constraints. Due to the complexity of the task we limit our model space to theories where the IMF has the shape prescribed by Eq. \ref{eq:1param_IMF} with a variable characteristic scale $\mu$, which is \emph{determined by the properties of the star-forming cloud}. For simplicity we further assume that $\mu$ can be approximately described as a power law in the form of
\be
\mu\propto \rho^{\gamma_{\rho}}\times T^{\gamma_{T}}\times \mach^{\gamma_{\mach}} \times R^{\gamma_{R}} \times Z^{\gamma_{Z}} \times B^{\gamma_{B}}
\label{eq:mu_powerlaw}
\ee
where $\rho$ is the density, $T$ is the temperature, $R$ is the size scale, $\mach$ is the turbulent Mach number, $Z$ is the metallicity and $B$ is the average magnetic field of the progenitor cloud. There are numerous examples of such IMF models, which have a somewhat fixed IMF shape and tie the characteristic mass to some property of the progenitor cloud, like Jeans mass \citep[e.g.][]{batebonell2005},  turbulent properties \citep[e.g.][]{hc08,core_imf,padoan_nordlund_2011_imf} or pressure \citep[e.g.][]{krumholz_stellar_mass_origin}.

\subsection{IMF constraints}\label{sec:constraints}
 
\begin{figure}
\begin {center}
\includegraphics[width=\linewidth]{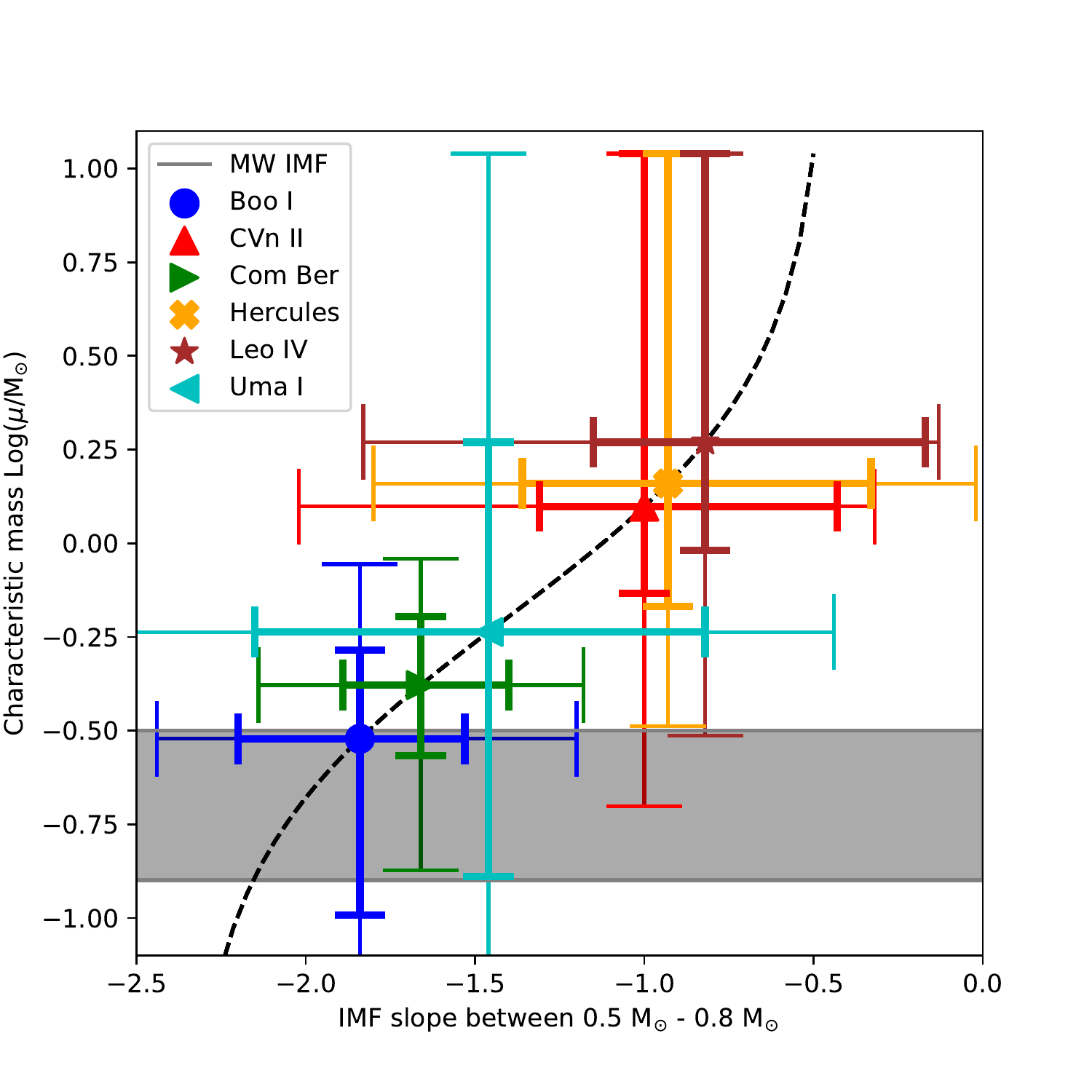}
\vspace{-1cm}
\caption{IMF slopes (between $0.5-0.8\,\msun$) and characteristic masses of UFD galaxies  based on the data of \protect\cite{Gennaro_2018_UFD_IMF1}. The errorbars correspond to 1$\sigma$ (thicker line) and 2$\sigma$ (thinner line) confidence intervals. The dashed line shows the mapping between the IMF slope and the $\mu$ characteristic mass from Fig. \ref{fig:mu_vs_slope}. The shaded region shows the possible values of the IMF characteristic mass in the MW. Of the 6 UFD galaxies, 3 have characteristic masses that are within $1\sigma$ of the MW values and all of them are within $2\sigma$. A characteristic mass that is roughly a factor of 2 higher than the MW value would be within $1\sigma$ for all galaxies, so we adopt that value as the average IMF shift for UFD galaxies.}
\label{fig:Gennaro2018_slopes}
\vspace{-0.5cm}
\end {center}
\end{figure}

 In this paper we investigate the effects of the following (proposed) observational constraints on the IMF model space (see Table \ref{tab:contraints} for quick summary):
 
\begin{enumerate}
\item \textbf{Universal IMF in the MW}: It has been fairly well established in the literature that the IMF in the MW is close to universal, regardless of the age and location of the stellar population \citep[see reviews of][]{imf_review,imf_universality, Hopkins_A_2018_IMF_obs_review}. Slight variation is possible in the characteristic mass on which we impose the conservative estimate of 0.2 dex based on Figure 3 of \citealt{imf_review}. Furthermore, based on resolved star counts the IMFs of old stellar populations have a similar or slightly more massive peak (see Figure \ref{fig:MN15_obs} and the references in the caption), still within the 0.2 dex limit\footnote{Note that dynamical evolution significantly alter the mass function of globular clusters leading to an apparent shift of the IMF peak to higher masses in studies that do not account for these effect \citep{Baumgardt_2008_GC_dynamical_evol,Kruijssen_2009_dynamical_evol}. After correcting for these biases one can recover a near-universal IMF in the MW for populations of all ages.}.
\item \textbf{MW-like IMF in dwarfs}: Dwarf galaxies like the LMC and SMC appear to have the same IMF as the MW despite different galactic metallicity, stellar mass and turbulent properties \citep[see review][]{imf_universality}. Note that the completeness limit of these studies is $>0.3\,\msun$ \citep[see][]{DaRio_2009_LMC_IMF, Gouliermis_2012_magellanic_clouds_IMF} so the peak of the IMF is not actually resolved, thus some variation is possible.
\item \textbf{Top-heavy IMF in ultra-faint dwarf galaxies}: Several recent observations have been extrapolated to imply top-heavy IMFs in ultra faint dwarf (UFD) galaxies \citep[see][]{Geha_2013_UFD_IMF,Gennaro_2018_UFD_IMF1,Gennaro_2018_UFD_IMF2} but there is no consensus in the field about these claims \citep[e.g.][]{Offner_2016_IMF_progress}. As Fig \ref{fig:Gennaro2018_slopes} shows, these results do not rule out a MW-like IMF with high confidence but are plausibly consistent with having a factor of 2 higher $\mu$ than in the MW. Due to this uncertainty, we explore the constraints arising from either having a top-heavy IMF or MW-like UFD IMF.
\item \textbf{Bottom-heavy in early-type galaxies}: There is growing indirect evidence suggesting that centers of early-type galaxies (ETGs) may have IMFs that are significantly more bottom heavy than the MW IMF \citep[e.g.][]{Conroy_vanDokkum_ellipticals,vanDokkum_conroy_2011,ConryVanDokkum_IMFvar_2012,Treu_galaxy_IMF_2010,Sonnenfeld_Treu_2015_IMF,Cappellari_IMF_var_2012}. A recent study by \cite{Conroy_2017_parametric_IMF} put the characteristic mass for one such galaxy below $0.1\,\msun$, about a factor of 3 smaller than the MW value.
\item \textbf{MW-like IMF in early-type galaxies}: Several recent studies using gravitational lensing have found ETGs to have mass-to-light ratios consistent with a MW-like IMF \citep[see][]{Collier_ETG_no_IMF_var_2018,Collier_2018_elliptical_galaxy_IMF}. This contradicts the results from studies using stellar population synthesis models. In this paper we will investigate the effects of both constraints.
\end{enumerate}

\subsection{Simulations}\label{sec:sim}

We utilize several simulated galaxies from the Feedback in  Realistic Environments (FIRE) project (\citealt{Hopkins2014_FIRE})\footnote{\url{http://fire.northwestern.edu}}. These galaxies have been presented in detail in \cite{Hopkins2017_FIRE2}, \cite{angles_alcazar_2017_fire_massive_bh} and \cite{Graus_2019_FIRE_dwarf} with one exception (\textbf{z5m12c}) that we will later discuss in \S~\ref{sec:ETG_galaxy} (also, see \S~\ref{sec:UFDgalaxy} on how we choose our UFD proxy). These are cosmological \myquote{zoom-in} simulations, which means that the simulation starts from a large cosmological box that is later rerun with increased resolution in areas of matter concentration (\myquote{zooms-in} on galaxies). The simulations proceed from $z>100$ to present day (except \textbf{z5m12c} and \textbf{h29\_HR}, see \S~\ref{sec:ETG_galaxy}). They are run using the GIZMO code \citealt{Hopkins2015_GIZMO})\footnote{\url{http://www.tapir.caltech.edu/~phopkins/Site/GIZMO.html}}, with the mesh-free Godunov ``MFM'' method for the hydrodynamics \citep{Hopkins2015_GIZMO}. Self-gravity is included with fully-adaptive force and hydrodynamic resolution. The simulations include detailed metallicity-dependent cooling physics from $T=10-10^{10}\,$K, including photo-ionization/recombination, thermal bremsstrahlung, Compton, photoelectric, metal line (following \citealt{Wiersma2009_cooling}), molecular, fine structure (following \citealt{CLOUDY}), dust collisional and cosmic ray processes, including both a meta-galactic UV background and each star in the simulation as a local source. The mass resolution for individual simulations is fixed and varies between $M_{\rm min}=250-56000\,\msun$ among our simulated galaxies (see Table \ref{tab:galaxies}).

The resolution of these cosmological simulations is not high enough to resolve the formation of individual stars ($M_{\rm min}\gg 0.01\,\msun$), instead star formation is approximated from resolved scales using a sink particle method. Gas is transformed into a sink particle if it:
\begin{enumerate}
\item Is locally self-gravitating
\item Is self-shielding
\item Is Jeans-unstable
\item Exceeds a minimum density threshold ($n>n_{\rm crit}$, see Table \ref{tab:galaxies})
\end{enumerate}
Such a sink particle is transformed into a \myquote{star cluster sink particle} on its dynamical time. Each of these represent a stellar population with the same formation properties (age, metallicity etc.) an thus the same IMF.

These \myquote{star cluster sink particles} provide feedback to the simulation via OB \&\ AGB mass-loss, SNe Ia \&\ II, and multi-wavelength photo-heating and radiation pressure; with inputs taken directly from stellar evolution models \citep{1Leitherer_1999_Starburst99}, assuming (in-code) a universal IMF \citep{Kroupa_IMF}\footnote{A major caveat of our analysis is that the feedback processes in the simulations assume a Kroupa IMF, so our post-processing neglects the potential feedback from a varying IMF and how it could enhance or suppress further IMF variation in a galaxy.}.

In this work, similar to \cite{guszejnov_imf_var}, we use cosmological simulations instead of present-day observations because they give us access to the entire star formation history of a galaxy. In a simulation we know the properties of star forming progenitor clouds \emph{at all times}, allowing us predict the IMF variation for the entire stellar population in a galaxy. 

\begin{table*}
	\centering
		\setlength\tabcolsep{3.0pt} 
		\begin{tabular}{|c|c|c|c|c|c|c|c|c|c|c|}
		\hline
		\bf Key & \bf Type & \bf Redshift & \bf $M_{\rm *}/\msun$ & \bf $M_{\rm DM}/\msun$ & $R_{\rm 1/2}/\mathrm{kpc}$ & \bf $M_{\rm min}/\msun$ & \bf $n_{\rm crit}/\mathrm{cm}^{-3}$ & $T_{\rm cloud}/\mathrm{K}$ & $\log(Z/Z_{\rm \sun})$ & $B/\mathrm{\mu G}$\\
        \hline
        \bf m12i & MW-like & $0$ & $6\times10^{10}$ & $10^{12}$ & $3.5$ & $7100$ & $10^3$ & 55  & 0.1 & 85\\ \hline
        \bf m11q & Dwarf, LMC-like & $0$ & $1.5\times 10^{9}$ & $10^{11}$ & $3.4$ & $7000$ & $10^3$ & 32   & -0.62 & 75\\ \hline
        \bf m10xf\_14706 & UFD, satellite & $0$ & $1.6\times 10^{5}$ & $10^{8.7}$ & $1.1$ & $4000$ & $10^3$ & 23  & -3.3 & N/A\\ \hline
        \bf h29\_HR & Early type & $2.5$ & $2\times 10^{11}$ & $10^{12.5}$ & 0.84 & $33000$ & $200$ & 92   & -0.15 & 256\\ \hline
        \bf z5m12c & Early type & $5$ & $3\times 10^{10}$ & $10^{12}$ & 3.3 & $56000$ & $200$ & 110   & -0.75 & N/A\\ \hline
		\end{tabular}
        \vspace{-0.1cm}
 \caption{Properties of simulated galaxies from the FIRE project, including stellar mass $M_{\rm *}$, target dark matter halo virial mass $M_{\rm DM}$ (at $z=0$), half-mass radius $R_{\rm 1/2}$, gas element mass resolution $M_{\rm min}$, critical density for sink particle creation $n_{\rm crit}$ and the galactic average temperature $T$, metallicity $Z$ and $B$ magnetic field (when available) for progenitor clouds. See Figure \ref{fig:gal_means} for further details.} 
 \label{tab:galaxies}\vspace{-0.5cm}
\end{table*}

\subsubsection{Finding a Proper Proxy for Massive Elliptical Galaxies}\label{sec:ETG_galaxy}

For our analysis we utilize simulated present-day galaxies where the \emph{at-formation} properties of all sink particles (star clusters) are available. We use these galaxies as proxies for the Milky Way (\textbf{m12i}), LMC and SMC (\textbf{m11q}) and UFDs (\textbf{m10q}). For early type galaxies we currently don't have access to such simulations so in this study we use three different proxies:
\begin{enumerate}
    \item \textbf{h29\_HR}: This is a simulated FIRE galaxy with additional black hole physics which lead to extreme starburst behavior, similar to what we expect in early type galaxies \citep[see simulation A2 in][for details]{angles_alcazar_2017_fire_massive_bh}. Unfortunately for these runs the at-formation properties of sinks were not saved. Re-running the simulation would have been very expensive, so instead we post-processed the about 150 snapshot files of the simulation, taking actively star-forming gas from each to approximate the distribution of progenitor cloud properties over cosmic time. In our previous study \citep{guszejnov_imf_var} we found that this approach provides a good approximation of the actual distribution. A major caveat with this proxy is that AGN feedback is poorly understood and thus it is not implemented in these simulations, despite the fact that it is believed to be one of the main mechanisms shutting off star formation.
    \item \textbf{z5m12c}: This run was originally conceived to study galaxy scaling relation in the era of reionization \citep[see][for details of the simulation setup]{Ma2018_reionization_FIRE}. It utilizes the same FIRE physics suite and the progenitor cloud properties are saved for all sink particles. Although this galaxy was simulated only to $z=5$, it is the progenitor of massive elliptical galaxy. The reason the simulation was not run further is due to the uncertainties in the physics that would quench star formation in such a galaxy. The stellar mass of \textbf{z5m12c} is only $3\times 10^{10}\,\msun$ at $z=5$, which is only a few percent of the mass it would attain by $z\sim 2$, the time from which most ETG IMF measurements are from. Nevertheless we can still use this simulated galaxy to look at the oldest population of stars in a ETGs.
    \item Both previous proxies have important caveats, so as a complementary approach we will approximate the properties of the progenitor clouds in ETGs using typical values for galaxies with extreme star formation (e.g. an Ultra Luminous Infrared Galaxy - ULIRG), see Table \ref{tab:canonical_ULIRG}. 
\end{enumerate}
Our 3 proxies essentially cover three possible ways to deal with the uncertain physics related to the quenching of star-formation in early type galaxies. With \textbf{h29\_HR} we neglect it and carry on with the simulation until $z=2.5$. In the case of \textbf{z5m12c} the simulation is stopped before AGN feedback could become important ($z=5$), so we are essentially assuming very strong, early quenching. Finally, by using the canonical ULIRG values we avoid the potential issues with the simulations, but these values are highly arbitrary.

\subsubsection{Finding a Proxy for Ultra Faint Dwarf Galaxies}\label{sec:UFDgalaxy}

To find an appropriate proxy for a UFD galaxy we use the simulations of \cite{Graus_2019_FIRE_dwarf}. With a combination of the {\tt Rockstar} halo finder \citep{Behroozi13}, and the Amiga Halo Finder \citep[\tt{AHF}]{Knollmann09} we identify halos and then select for galaxies with a low stellar mass ($<10^6\,\msun$). In \textbf{m10xf} there are over 150 such low-mass, from these we restrict our study to those whose stellar population is well-resolved (>25 \myquote{star} sink particles), which essentially sets the lower bound to our galaxy masses as $10^5\,\msun$. This leaves 8 galaxies, for our study we choose \textbf{m10xf\_14706}, the one with the lowest stellar mass ($1.5\times 10^5\,\msun$). Note that picking a different galaxy from this group does not change the qualitative results of this study.

\begin{table}
	\centering
		\begin{tabular}{|c|c|}
		\hline
        \multicolumn{2}{|c|}{\bf ULIRG-like cloud star-forming clouds} \\
        \hline
		Density &  $2\times 10^5\,\mathrm{cm}^{-3}$\\
        \hline
		Temperature &  $75\,\mathrm{K}$\\
        \hline
		Turbulent dispersion ($\sigma_{T}$) &  $10\,\mathrm{km}/\mathrm{s}$\\
        \hline
		Metallicity ($\log(Z/Z_{\rm \sun})$ &  $0.5$ \\
        \hline
		Magnetic field &  100$\,\mathrm{\mu G}$ \\
        \hline
		\end{tabular}
        \vspace{-0.1cm}
 \caption{ULIRG-like values assumed for the star-forming clouds in early-type galaxies that we use for testing IMF constraints.} 
 \label{tab:canonical_ULIRG}\vspace{-0.5cm}
\end{table}


\begin{table*}
	\centering
		\begin{tabular}{|c|cc|c|c|}
		\hline
		\textbf{Environment} & \multicolumn{2}{c|}{\textbf{Constraint}}  & \textbf{Reference} & \textbf{Proxy} \\
        \hline
        \multirow{2}{*}{Milky Way} & Universal IMF & $\sigma_{\mu,\mathrm MW}<0.2$  & \cite{imf_universality} & \multirow{2}{*}{\textbf{m12i}}  \\
        \cline{2-3} 
        & Universal IMF & $0<\Delta\mu_{\rm MW, z>3}<0.2$ & \cite{imf_review} and Figure \ref{fig:MN15_obs} &  \\
        \hline
        Dwarf Galaxies & MW-like IMF & $\left|\Delta\mu_{\rm DG}\right|<0.2$  & \cite{imf_universality} & \textbf{m11q}  \\
        \hline
        \multirow{2}{*}{UFD Galaxies} & Top-heavy IMF & $\left|\Delta\mu_{\rm UFD}-\log 2\right|<0.2$  & \cite{Gennaro_2018_UFD_IMF1} & \multirow{2}{*}{\textbf{m10xf\_14706}} \\
       \cline{2-3}
        & MW-like IMF & $\left|\Delta\mu_{\rm UFD}\right|<0.2$ & \cite{Offner_2016_IMF_progress} &   \\
        \hline
        \multirow{6}{*}{Early Type Galaxies} & \multirow{3}{*}{Bottom-heavy IMF} & \multirow{3}{*}{$\left|\Delta\mu_{\rm ETG}-\log 1/3\right|<0.2$}  & \multirow{3}{*}{\cite{Conroy_2017_parametric_IMF}} & \textbf{z5m12c}  \\
        \cline{5-5}
        & & & & \textbf{h29\_HR} \\
        \cline{5-5}
        & & & & ULIRG values  \\
        \cline{2-5}
        & \multirow{3}{*}{MW-like IMF}  & \multirow{3}{*}{$\left|\Delta\mu_{\rm ETG}\right|<0.2$}  & \multirow{3}{*}{\cite{Collier_2018_elliptical_galaxy_IMF}} & \textbf{z5m12c}  \\
        \cline{5-5}
        & & & & \textbf{h29\_HR}  \\
        \cline{5-5}
        & & & & ULIRG values \\
		\hline
		\end{tabular}
        \vspace{-0.1cm}
 \caption{Summary of observations constraints on the IMF in various environments as well as the proxies (mostly simulated galaxies from FIRE project) we use to estimate the constraints they put on IMF models. Here $\Delta\mu=\log \mu/\mu_{\rm MW}$ is the amount (in dex) the IMF characteristic mass is shifted in different environments, while $\sigma_{\mu,\rm MW}$ is the standard variation of $\log\mu$ in the Milky Way.}
 \label{tab:contraints}
 \vspace{-0.5cm}
\end{table*}

\vspace{-0.5cm}
\subsection{From Parent Cloud to IMF Properties}\label{sec:imf_models}
\label{sec:methods.imf.models}

\begin{figure}
\begin {center}
\includegraphics[width=\linewidth]{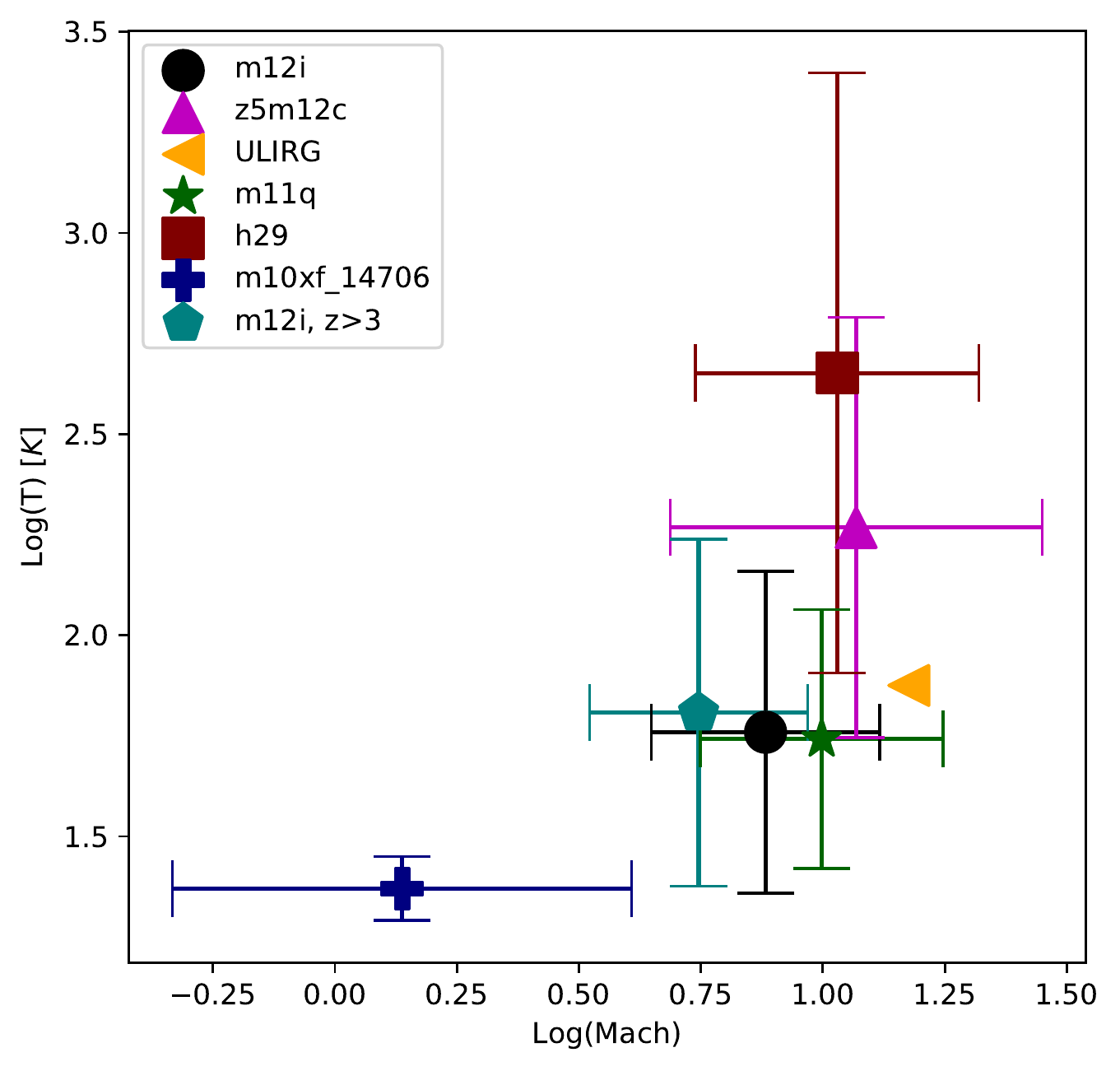}\\
\includegraphics[width=\linewidth]{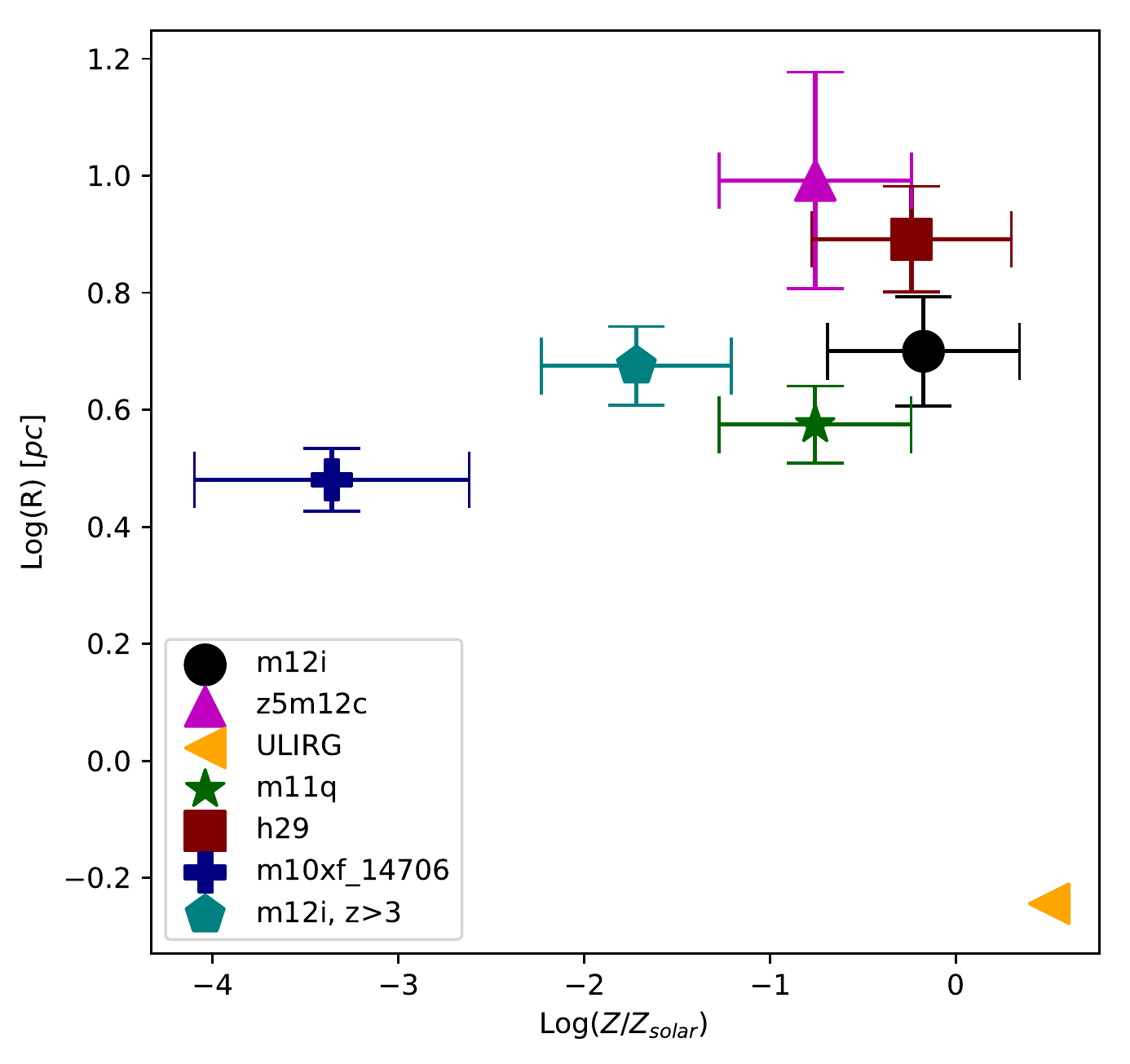}
\vspace{-0.6cm}
\caption{Galaxy-average properties of progenitor clouds (\textit{Top}: turbulent Mach number vs temperature, \textit{Bottom:} metallicity vs cloud size) in the simulated galaxies (see Table \ref{tab:galaxies} for more details on the individual runs). The mean values are galaxy-scale averages of the logarithmic quantity weighted by stellar mass, while the errorbars show the corresponding $1\sigma$ scatter. There is significant variation not only between the different galaxies but also within individual galaxies. The average properties of star-forming clouds evolve significantly during a galaxy's lifetime, this is why this scatter is much larger than the observed scatter in \emph{present-day} star-forming clouds.}
\label{fig:gal_means}
\vspace{-0.5cm}
\end {center}
\end{figure}

\begin{figure*}
\begin {center}
\includegraphics[width=0.49\linewidth]{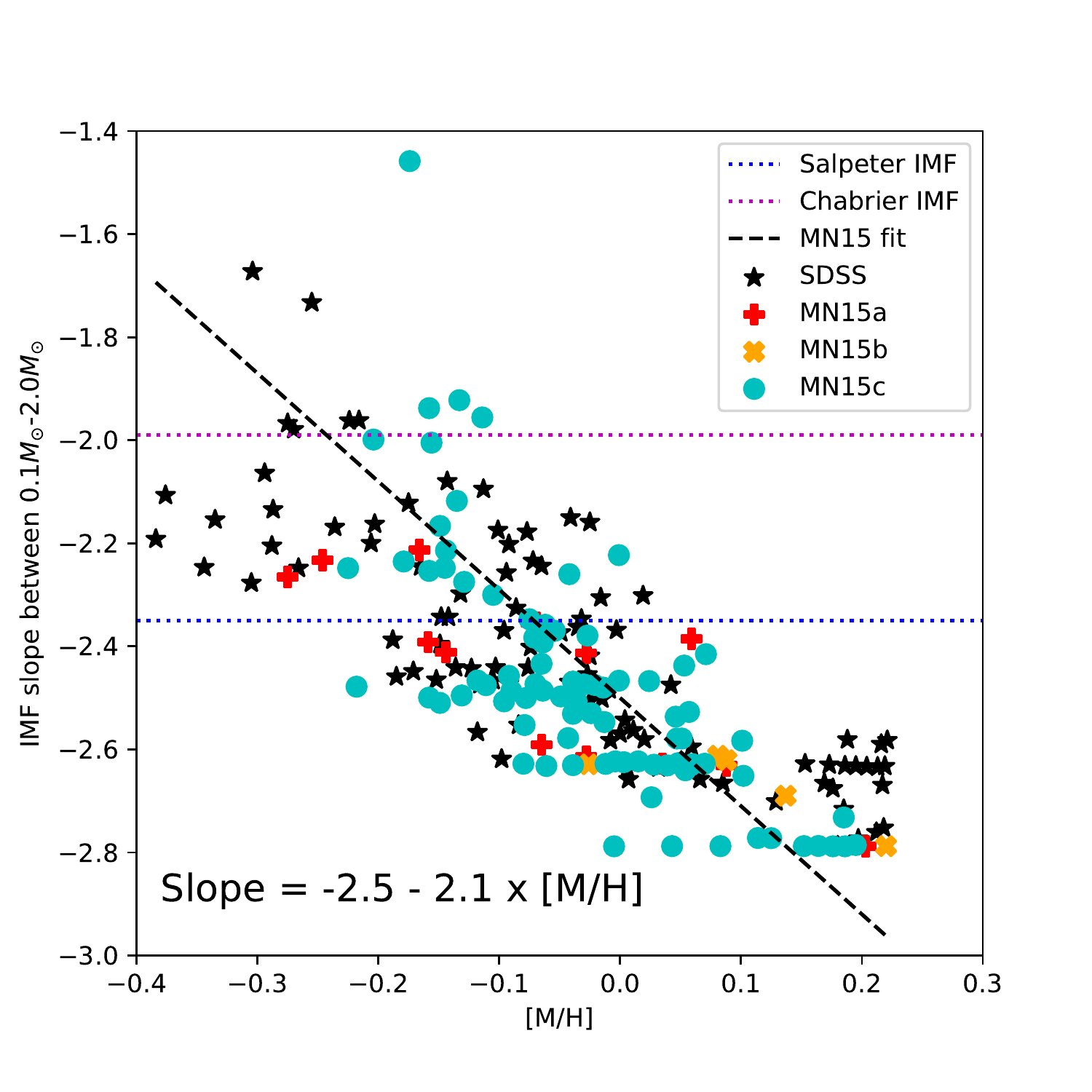}
\includegraphics[width=0.49\linewidth]{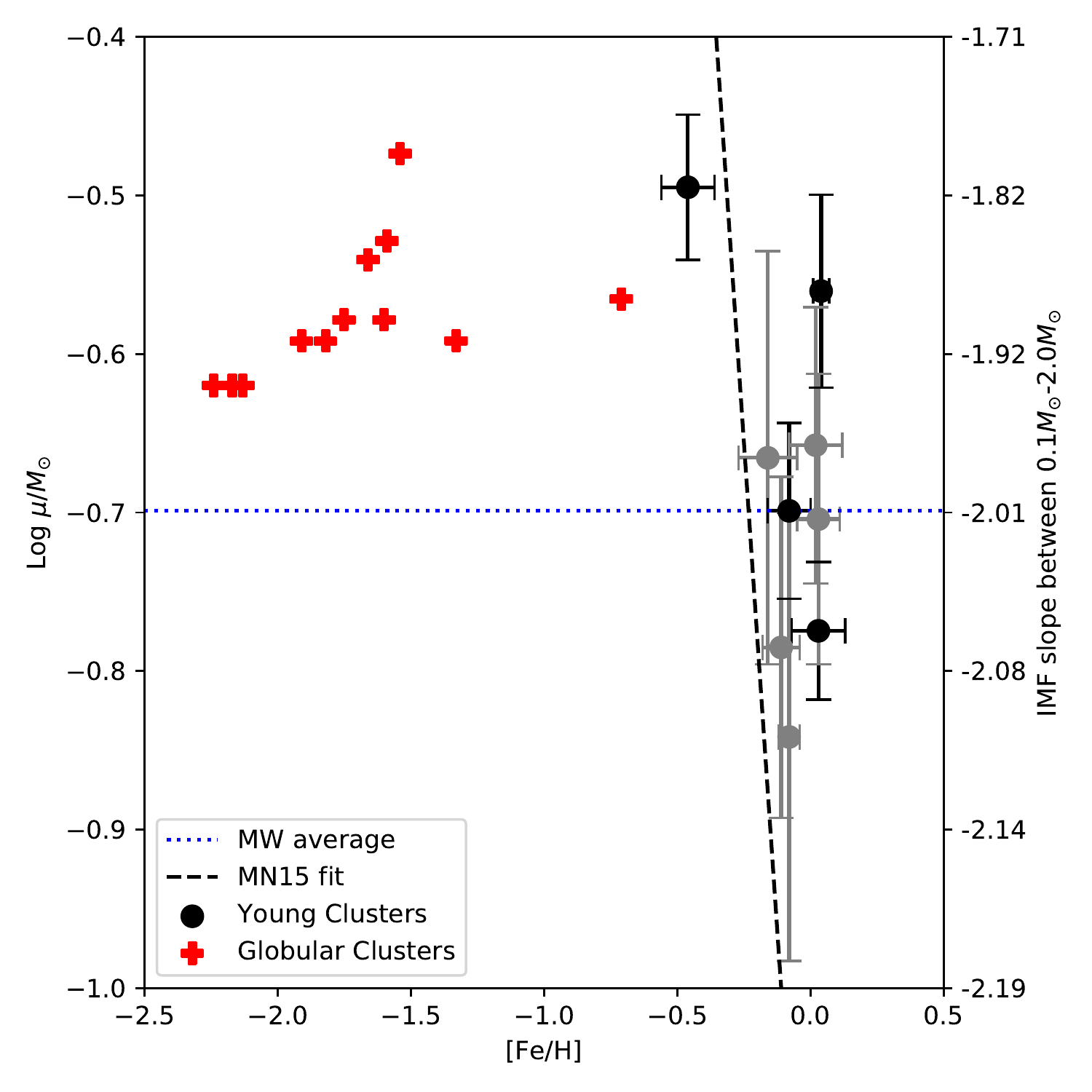}
\vspace{-0.5cm}
\caption{\textbf{Left:} Galaxy-averaged single-power-law IMF slopes\protect\footnotemark \,inferred by stellar population synthesis models from measured spectra by \citealt{navarro_2015a,navarro_2015_imf_relic,navarro_2015_imf_relation} plotted versus galactic metallicity (see Figure 2 of \citealt{navarro_2015_imf_relation} for uncertainties). There is a significant correlation between the inferred IMF slope and the galactic metallicity well-described by a linear fit (black, dashed line). \textbf{Right:}  IMF peaks inferred in the Milky Way from resolved star counts for young star forming regions \citep[][in black, the ones with large errorbars are plotted in grey]{Luhman_2007_ChaI_starcount_IMF,Oliveira_2009_NGC6611_starcount_IMF,Sung_Bessel_2010_NGC2264_starcount_IMF,Bayo_2011_lambdaOri_starcount_IMF,Lodieu_2011_IC4665_starcount_IMF,Lodieu_2012_alphaPer_starcount_IMF,Lodieu_2012_Pleiades_starcount_IMF,PenaRamirez_2012_SigmaOrionis_starcount_IMF,AlvesdeOliveira_2013_IC348_starcount_IMF} for old stellar populations in globular clusters \citep[][in red]{Paresce_deMarchi_2000_GC_IMF} as well as the IMF peak-metallicity relation inferred from \citealt{navarro_2015_imf_relation}. It is clear that the resolved star counts in the MW rule out an IMF that is solely determined by metallicity as it was also shown by \citealt{Villaume_2017_GC_IMF_var_SPS}.}
\label{fig:MN15_obs}
\vspace{-0.5cm}
\end {center}
\end{figure*}

Because the simulations resolve down to cloud scales, but no further, we treat each star-forming gas element as an independent ``parent cloud'', which sets the initial conditions for its own detailed IMF model (in accordance with the IMF models we investigate). Specifically, whenever a sink particle is spawned, we record all properties of the parent gas element from which it formed, and use these in post-processing to predict the IMF for the stellar population it spawns. 
From this point we infer the IMF characteristic mass $\mu$ from the initial conditions of the parent clouds that form stars in the simulations \citep[see][ for an example of how GMC properties could be mapped to the IMF]{guszejnov_feedback_necessity}. While we investigate the entire model space described by Eq. \ref{eq:mu_powerlaw} we give special attention to the following classes of models that are common in the literature (summarized in Table~\ref{tab:SF_models}):
\begin{itemize}
  \item \textbf{Jeans mass models:} Gas clouds collapse primarily through the Jeans instability. This model assumes that the initial Jeans mass of the progenitor cloud sets the characteristic mass of the stars it spawns (e.g. \citealt{batebonell2005}), so
	\be
	\mu_{\rm Jeans}\propto M_{\rm Jean}= \frac{\pi c_s^3}{6 G^{3/2}\rho^{1/2}}.
	\ee
    Note that the models may still assume sub-fragmentation to smaller scales, but the key assumption (for our purposes) is simply that the turnover mass is proportional to the parent cloud Jeans mass.
	
	\item \textbf{Opacity limited equation of state (EOS) models:} As clouds become denser they reach the point where they become opaque to their own cooling radiation, leading to a transition from isothermal to adiabatic behavior, suppressing fragmentation at the Jeans mass corresponding to this critical volume density $\rho_{\rm crit}$ (e.g. \citealt{LowLyndenBell1976,Whitworth98a,Larson2005,Glover_EQS_lowgamma_ref, Jappsen_EQS_ref, Masunaga_EQS_highgamma_ref}). Motivated by radiation transfer simulations like \cite{Bate_2009_rad_importance} we also investigated the case where the transition occurs at a critical surface density $\Sigma_{\rm crit}$. The resulting characteristic masses are:
	\begin{eqnarray}
	\mu_{\rm EOS,\rho}\sim \frac{\pi c_s^3}{6 G^{3/2}\rho_{\rm crit}^{1/2}},&
	\mu_{\rm EOS,\Sigma}\sim \frac{c_s^4}{G^2 \Sigma_{\rm crit}},
	\end{eqnarray}
	where $\rho_{\rm crit}$ and $\Sigma_{\rm crit}$ are the critical densities for the isothermal-adiabatic transition.
	
	\item \textbf{Turbulent/sonic mass models:} Several analytical theories derive the core mass function (CMF) and the IMF from the properties of the turbulent medium, in which they form (e.g. \citealt{Padoan_Nordlund_2002_IMF,hc08,core_imf,hc_2013}). In these models, both the CMF and IMF peaks are set by the ``sonic mass'' $M_{\rm sonic}$, namely the turbulent Jeans or Bonner-Ebert mass at the sonic scale ($R_{\rm sonic}$) below-which the turbulence becomes sub-sonic and therefore fails to generate large density fluctuations (which seed fragmentation). The various theories give slightly different answers to this critical mass, in this paper we will use the definitions of \cite{core_imf} and \cite{hc_2013}, which give:
	\begin{eqnarray}
	\mu_{\rm sonic} \sim M_{\rm sonic}\sim \frac{2 c_s^2 R_{\rm sonic}}{G}\sim M_{\rm Jeans}/\mach\\
	\mu_{\rm HC2013} \sim M_{\rm Jeans}/\mach^{2} \sim M_{\rm sonic}/\mach,
	\end{eqnarray}
  where $R_{\rm sonic}$ is defined through the linewidth-size relation
  \be 
  \sigma^2_{\rm turb}(\lambda)=c_s^2 \frac{\lambda}{R_{\rm sonic}}.
  \ee
 In our simulations $\sigma_{\rm turb}^2$ is estimated for a progenitor cloud (sink particle) by measuring the velocity dispersion (after subtracting the mean shear) between neighboring gas particles in a sphere of radius $\lambda$ (taken to be that which encloses the nearest $\sim 32$ gas neighbours). 
	
\item \textbf{Protostellar feedback models:} Although there are a number of ways newly-formed stars can regulate star formation, most studies have concluded that at the scale of the IMF peak (early protostellar collapse of $\sim 0.1\,\msun$ clouds) the most important self-regulation mechanism is radiative feedback from protostellar accretion \citep{Bate_2009_rad_importance, krumholz_stellar_mass_origin, guszejnov_feedback_necessity}. This sets a unique mass and spatial scale within which the protostellar heating has raised the temperature to make the core Jeans-stable, suppressing fragmentation. The resulting critical mass is 
\be
\mu_{\rm K11}\sim 0.15\left(\frac{P/k_B}{10^{6}\,\mathrm{K/cm^3}}\right)^{-1/18}\solarmass
\ee
where $P$ is the pressure of the gas. There are several other formulas in the literature (e.g. \citealt{Bate_2009_rad_importance}); the differences are due to the detailed uncertainties in the treatment of radiation. However, for our purposes, they give {\em nearly identical} results, so we will focus on the model from \citet{krumholz_stellar_mass_origin}.

\item \textbf{Metallicity dependent IMF models:} Some SPS analyses of early-type galaxies have been empirically fit by assuming a trend of increasingly shallow IMF slopes with decreasing metallicity (see \citealt{navarro_2015_imf_relation} and Figure \ref{fig:MN15_obs}). This phenomenological model sets the slope of the IMF (in the $>0.6\,\solarmass$ range) as:
\be 
\mathrm{Slope}[0.6\,\msun<M]=-2.2-3.1\times \mathrm{[M/H]},
\ee
where M/H is the logarithm of the mass-weighted total metallicity relative to the solar value. Note that the actual measurements are only sensitive to the IMF in the $0.1\,\msun-2.0\,\msun$ regime and the above relation was derived in \citealt{navarro_2015_imf_relation} by assuming a two-part IMF with fixed parameters below $0.6\,\msun$. To preserve generality, we use instead the single-power-law IMF fit from that work which yields:
\be 
\mathrm{Slope}[0.1\,\msun<M<1.5\,\msun]=-1.5-2.1\times \mathrm{[M/H]}.
\ee
Using the one-parameter model from Eq. \ref{eq:1param_IMF}, we can convert the metallicity-slope relation into the $\mu$-metallicity relation of
\be 
\log(\mu/\msun)= -1.3 -2.4\times \mathrm{[M/H]} + \ord\left(\mathrm{[M/H]}^2\right).
\ee
As shown in Figure \ref{fig:MN15_obs} this phenomenological model provides a good fit for the inferred extragalactic IMFs but drastically overpredicts the variations for old stellar populations within the MW.
\end{itemize}

\footnotetext{The measurements of \citealt{navarro_2015a,navarro_2015_imf_relic,navarro_2015_imf_relation} are sensitive to the $F_{05}$ dwarf-to-giant ratio in the $0.1\,\msun-2.0\,\msun$ range. The IMF slope is inferred by assuming that it is well-described by a single power-law in this regime.}

\section{Results and Discussion}\label{sec:results}

Using our simulation proxies we can calculate the shifts and variations of the IMF for the previously mentioned models. Table \ref{tab:SF_models} shows although some models can come close to reproducing the claimed UFD and ETG IMF variations (e.g. \citealt{hc_2013}), these drastically violate IMF universality within the MW. We find that \emph{none of the current models in the literature can satisfy all constraints}, so we extend our search to generic models following Equation \ref{eq:mu_powerlaw}.

\begin{table*}
	\centering
	\setlength\tabcolsep{2.0pt} 
		\begin{tabular}{|c|c|c|c|c|c|c|c|c||c|c|}
		\hline
		\multirow{3}{*}{\bf IMF Model} & \multirow{3}{*}{\bf IMF scale ($\mu$)} & \multirow{3}{*}{\bf Reference} & \multirow{3}{*}{\bf Key} & \multicolumn{7}{c|}{\bf IMF variation/shift [dex]} \\
		 &  & &  & \multicolumn{5}{c||}{sink particle average} & \multicolumn{2}{c|}{bulk properties} \\
        & & & & $\sigma_{\mu,\mathrm MW}$ & $\Delta\mu_{\rm MW, z>3}$ & $\Delta\mu_{\rm DG}$ & $\Delta\mu_{\rm UFD}$ & $\Delta\mu_{\rm HZ}$ & $\Delta\mu_{\rm H29}$ & $\Delta\mu_{\rm ULIRG}$ \\
        \hline
		Jeans Mass & $\propto T^{3/2}\rho^{-1/2}$ & \citealt{batebonell2005} & M\_Jeans & \red{0.71} & \green{0.10} & \green{-0.03} & \red{-0.88} & \red{0.65} & \red{1.24} & \green{-0.99} \\
        \hline
		\multirow{2}{*}{Turbulent/Sonic Mass} & $\propto T^{3/2}\rho^{-1/2}\mach^{-1}$  & \citealt{core_imf} & M\_Sonic  & \red{0.96} & \orange{0.30} & \green{0.05} & \red{-0.13} & \red{0.47} & \red{1.16} & \green{-1.28} \\
        \cline{2-11}
        & $\propto T^{3/2}\rho^{-1/2}\mach^{-2}$  & \citealt{hc_2013} & HC\_2013  & \red{1.98} & \orange{0.37} & \green{0.12} & \orange{0.61} & \red{0.28} & \red{1.01} & \green{-1.57} \\
        \hline
		Opacity-limited, $\rho_{\rm crit}$ & $\propto T^{3/2}$ & \citealt{Jappsen_EQS_ref} & EOS\_Rho  & \red{0.60} & \green{0.07} & \green{-0.06} & \red{-0.58} & \red{0.76} & \red{1.27} & \red{0.17} \\
        \hline
		Opacity-limited, $\Sigma_{\rm crit}$ & $\propto T^{2}$ & \citealt{Bate_2009_rad_importance} & EOS\_Sigma  & \red{0.80} & \green{0.10} & \green{-0.07} & \red{-0.78} & \red{1.02} & \red{1.69} & \red{0.23} \\
        \hline
		Protostellar Heating & $\propto \left(\rho T\right)^{-1/18} $  & \citealt{krumholz_stellar_mass_origin} & Heating\_K11  & \green{0.03} & \green{0.0001} & \green{0.005} & \orange{-0.01} & \orange{-0.04} & \orange{-0.05} & \orange{-0.14} \\
\hline
		Metallicity dependent & $\propto [\mathrm{M/H}]^{-2.4} $  & \citealt{navarro_2015_imf_relation} & MN\_2015  & \red{1.24} & \red{3.7} & \red{1.73} & \red{7.64} & \red{1.40} & \red{0.39} & \green{-1.62} \\
\hline
		\end{tabular}
        \vspace{-0.1cm}
 \caption{The different IMF models compared in this paper, each with the predicted scaling of the IMF characteristic mass $\mu$ with initial parent cloud properties (\S~\ref{sec:methods.imf.models}), reference and the predicted IMF variations/shifts. $\sigma_{\mu,\mathrm MW}$ is the galaxy-wide scatter of the $\mu$ characteristic mass in a simulated MW-like galaxy (\textbf{m12i}), while $\Delta\mu_{\rm MW, z>3}$, $\Delta\mu_{\rm DG}$, $\Delta\mu_{\rm UFD}$, $\Delta\mu_{\rm HZ}$  are the amount $\log\mu$ \myquote{shifts} in the simulated old MW populations, dwarf galaxy (\textbf{m11q}), UFD analogue (\textbf{m10xf\_14706}), high z early-type galaxy (\textbf{z5m12c}) and early-type galaxy (\textbf{h29\_HR}). $\Delta\mu_{\rm ULIRG}$ (last column) shows the shift we get if we approximate the properties of the progenitor clouds in an early-type galaxy with those of a ULIRG (see Table \ref{tab:canonical_ULIRG}). The values for these shifts/variations are color coded based on how close they are to satisfying the constraints of Table \ref{tab:contraints}. We use the following scale: \green{green} if they satisfy the constraint, \orange{orange} if they fail to do so are within 0.2 dex (one MW IMF scatter) and \red{red} if they grossly violate the constrain (>0.2 dex difference). The only model that can satisfy MW universality is the protostellar heating based IMF model, which produces almost zero IMF variations in all cases.} 
 \label{tab:SF_models}\vspace{-0.5cm}
\end{table*}

\begin{figure}
\begin {center}
\includegraphics[width=\linewidth]{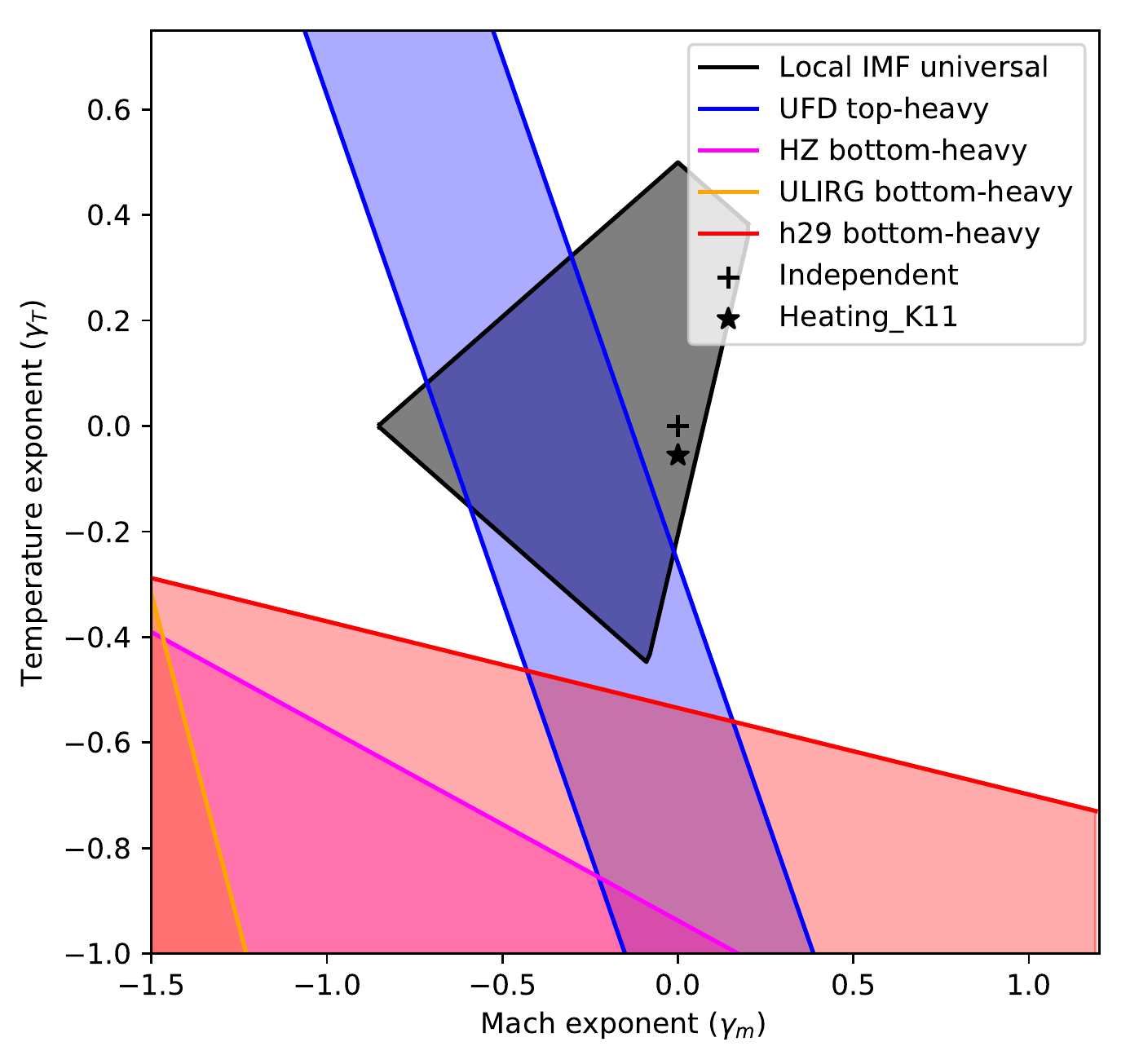}\\
\includegraphics[width=\linewidth]{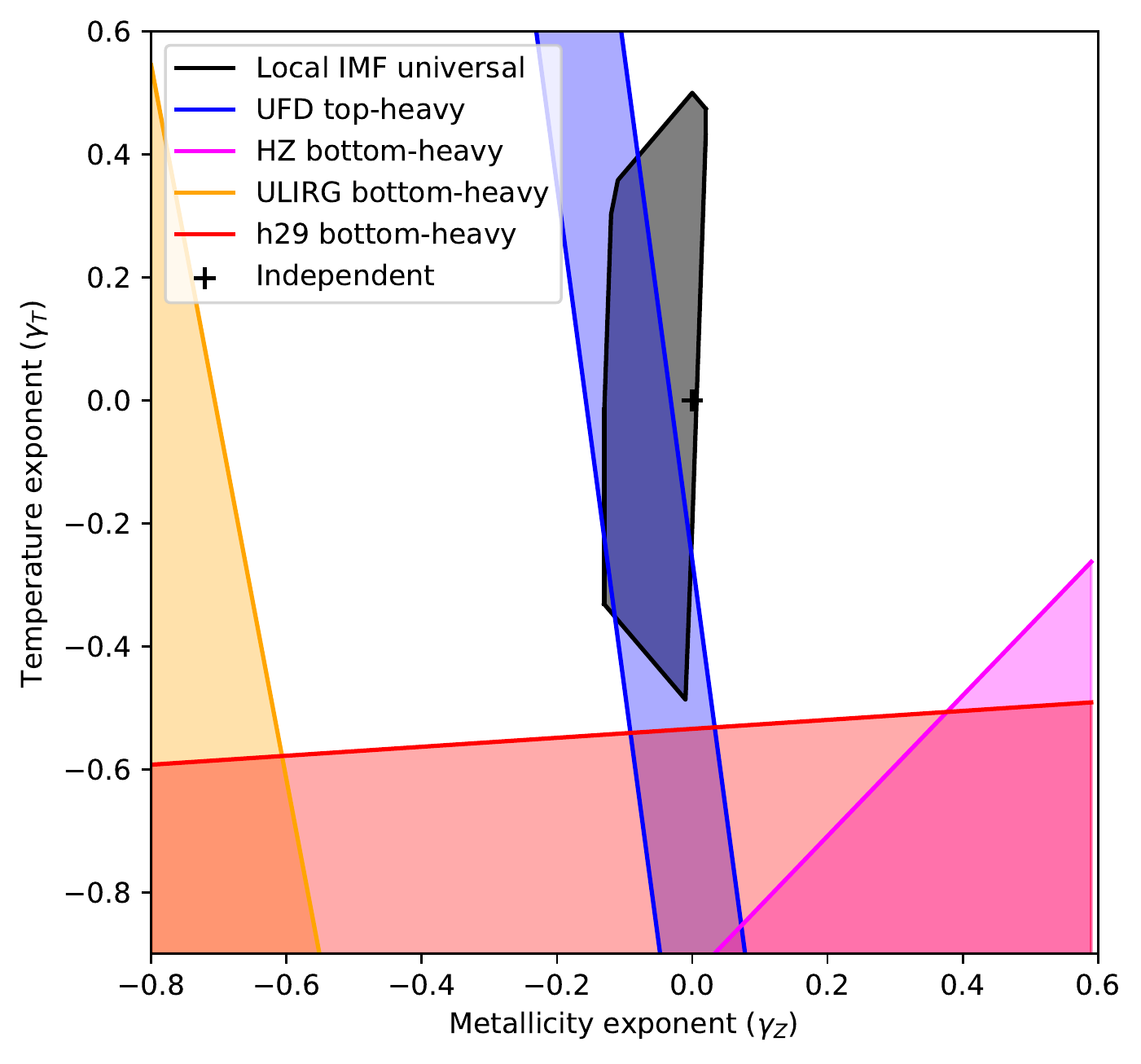}
\vspace{-0.6cm}
\caption{Power-law exponents for the Mach-number and temperature (left) and metallicity and temperature (right) in Equation \ref{eq:mu_powerlaw}. The shaded regions show the exponents that satisfy IMF universality (grey region, see Figure \ref{fig:exponents_universal} for details), reproduce the inferred top-heavy IMF for ultra faint dwarfs (blue region) and the bottom-heavy IMF for early type galaxies (red, magenta and orange regions, see Table \ref{tab:contraints} for details on constraints and Table \ref{tab:galaxies} on the simulated galaxies). The symbols represent IMF models from the literature (see Table \ref{tab:SF_models}). There is clearly no overlap between the different types of constraints, thus there is no IMF model in this form that can satisfy all constraints.}
\label{fig:exponents_both}
\vspace{-0.5cm}
\end {center}
\end{figure}

Using the variations in progenitor cloud properties in the MW-like galaxy of \textbf{m12i} we can identify the IMF models (the exponents for Eq. \ref{eq:mu_powerlaw}) that would satisfy IMF universality in the MW (this exercise is worked out in detail by \cite{guszejnov_imf_var}). With our additional constraints for old MW populations, dwarf galaxies, UFDs and ETGs from \S~\ref{sec:constraints} we can further restrict the model space, see Appendix \ref{sec:exponents}. We investigate both models from the literature (see Table \ref{tab:SF_models} for results) and generic models following Equation \ref{eq:mu_powerlaw} (see Figure \ref{fig:exponents_both} for an example). In general, we can draw the following conclusion:
\begin{itemize}
\item The large difference in average cloud metallicity in the older ($z>3$) stellar population in our MW-like galaxies (\textbf{m12i},see Figure \ref{fig:gal_means}) compared to average and the significant scatter in metallicity in the latter, only allows the IMF characteristic mass to have a weak metallicity dependence (see Figure \ref{fig:exponents_both}).
\item There is little-to-no overlap between the regions that satisfy local IMF universality and those that reproduce the observed IMF shifts in ETGs (see Figure \ref{fig:exponents_both} and Figure \ref{fig:exponents_vars}).
\item MW universality strongly rules out most IMF models in the literature, including the turbulent/sonic mass models \citep[see][for a detailed analysis]{guszejnov_imf_var}.
\item There exists a significant region of the model space that satisfies the assumption that the IMF is near universal in all types of galaxies ($<0.2\,\mathrm{dex}$ scatter in the galactic mean $\mu$). An example of such \myquote{weakly varying IMF} models is the protostellar heating model of \citealt{krumholz_stellar_mass_origin} (Heating\_K11).
\end{itemize}

None of the models detailed in \S~\ref{sec:imf_models} can reproduce the IMF variations which have been claimed for either ETGs or UFDs without grossly violating constraints from the local measurements with resolved star counts. Meanwhile the only model that reproduces the MW observations (the \textit{Protostellar Heating} model) predicts essentially zero IMF variations in almost all environments. It is therefore natural to ask whether there even exists a model that can reproduce both the claimed variations and the near-universality in the MW. Progenitor clouds have essentially 6 (nearly) independent properties: size, density, temperature, Mach number, magnetic field strength and metallicity\footnote{In the simulations we used there is some correlation between these quantities, but we neglect them here for simplicity. Taking them into account does not change the result significantly.}. We are looking for the exponents corresponding to these quantities in Eq. \ref{eq:mu_powerlaw}. If these quantities are independent then MW universality allows us to restrict the space we search to a 6D rectangle whose sides are described by: $\left|\gamma_X \Delta\log X_{\rm MW}\right|< \Delta\log \mu_{\rm MW}$, where $\Delta \log X_{\rm MW}$ is how many orders of magnitude of scatter quantity $X$ has in our proxy for the MW (\textbf{m12i}), while $\Delta \log \mu_{\rm MW}$ is the maximum allowed scatter in the MW IMF peak (0.2 dex, see \S~\ref{sec:constraints}). Within this region we use a Monte-Carlo search to find a set of exponents that would satisfy all criteria. We find the following:
\begin{enumerate}
    \item There is a significant volume in the model space that satisfies local IMF universality and produces a top-heavy IMF UFD galaxies.
    \item In case of our ETG proxy where we followed the galaxy evolution to $z\sim 2.5$ without AGN effects we find that there is no IMF model in the shape of Equation \ref{eq:mu_powerlaw} that can satisfy local IMF universality and produce the claimed bottom-heavy IMF.
    \item There exist a small volume in the model space that seems to satisfy local IMF universality and reproduce the observed IMF variations in both UFDs and ETGs (either \textbf{z5m12c} or ULIRG values, not both). These models however do not correspond to any known physical mechanism (e.g. $\mu\propto R^{-3/2}$). Furthermore they all utilize the fact that the cloud sizes in \textbf{z5m12c} and in the ULIRG values are significantly different from the values in \textbf{m12i}. Note that the mass resolution and the critical density of the simulations (see Table \ref{tab:galaxies}) set a size scale that appears in the sink particle sizes and thus in the progenitor cloud sizes (essentially the size scale where the simulation replaces gas clouds with sink particles). To verify these models we use a lower mass resolution version of \textbf{m12i} ($\Delta m=56000\,\msun$, like in \citealt{guszejnov_imf_var}), which clearly rules out all of these models. This means that \emph{there is no generic model that satisfies all constraints}.
    \item If we relax the claimed variations in early-type galaxies (e.g. a factor 2 shift instead of 3) we find that a significant volume of the model space can produce appropriate bottom-heavy IMFs for both simulation proxies (\textbf{z5m12c} and \textbf{h29\_HR}) as well as satisfying local IMF universality and producing a top-heavy IMF for UFDs. Still, these models correspond to no known physical mechanism (e.g. $T^{-1/4}R^{-1/4}$).
\end{enumerate}

\section{Conclusions}\label{sec:consclusions}

In this paper we used different types of simulated galaxies to infer what constraints different observational claims impose on theoretical IMF models. We mainly focused on three common claims from the literature: 1) that the IMF in the MW and nearby dwarf galaxies is nearly universal, 2) that the IMF in early-type galaxies is \myquote{bottom-heavy} and 3) that the IMF in ultra-faint dwarf galaxies is \myquote{top-heavy}. We found that the current models in the literature either fail to reproduce the observed IMF variations or violate IMF universality in the MW.

We also investigated generic IMF models where the IMF characteristic mass is a power-law of progenitor cloud properties. Despite the high dimensionality of the model space, we find that no model where the turnover mass is an arbitrary power-law function of a combination of cloud temperature/density/size/metallicity/velocity dispersion/magnetic field can reproduce the claimed IMF variation in ellipticals or dwarfs without severely violating observational constraints in the Milky Way.

One possibility is that the characteristic mass of the IMF is set by a yet unknown physical mechanism. Another, more likely scenario is that the magnitude of of IMF variations in ETGs are overestimated in stellar population synthesis models. This would further explain why non-SPS based methods (e.g. gravitational lensing, see \citealt{Collier_ETG_no_IMF_var_2018}) appear to contradict SPS-based observations. There are a several possible reasons for such a bias, most of them coming from the inherent uncertainties of extrapolating stellar atmosphere models to extreme metallicities. We find that relaxing the claimed variations greatly increases the number of possible models.
 
\acknowledgments
Support for DG and PFH was provided by an Alfred P. Sloan Research Fellowship, NSF Collaborative Research Grant \#1715847 and CAREER grant \#1455342, and NASA grants NNX15AT06G, JPL 1589742, 17-ATP17-0214. Numerical calculations were run on the Caltech compute cluster ``Wheeler,'' allocations from XSEDE TG-AST130039 and PRAC NSF.1713353 supported by the NSF, and NASA HEC SMD-16-7592.
This work used computational resources of the University of Texas at Austin and the Texas Advanced Computing Center (TACC; http://www.tacc.utexas.edu), the NASA Advanced Supercomputing (NAS) Division and the NASA Center for Climate Simulation (NCCS).
DG and ASG were supported by the Harlan J. Smith McDonald Observatory Postdoctoral Fellowship.
We would like to thank Stella Offner, Daniel Angl\'es-Alc\'azar and Alexa Villaume for their help and comments.

\vspace{0.25cm}
\bibliographystyle{mnras}
\bibliography{bibliography}

\appendix

\section{Effects of individual constraints on the allowed IMF exponents}\label{sec:exponents}
\begin{figure*}
\begin {center}
\includegraphics[width=0.33\linewidth]{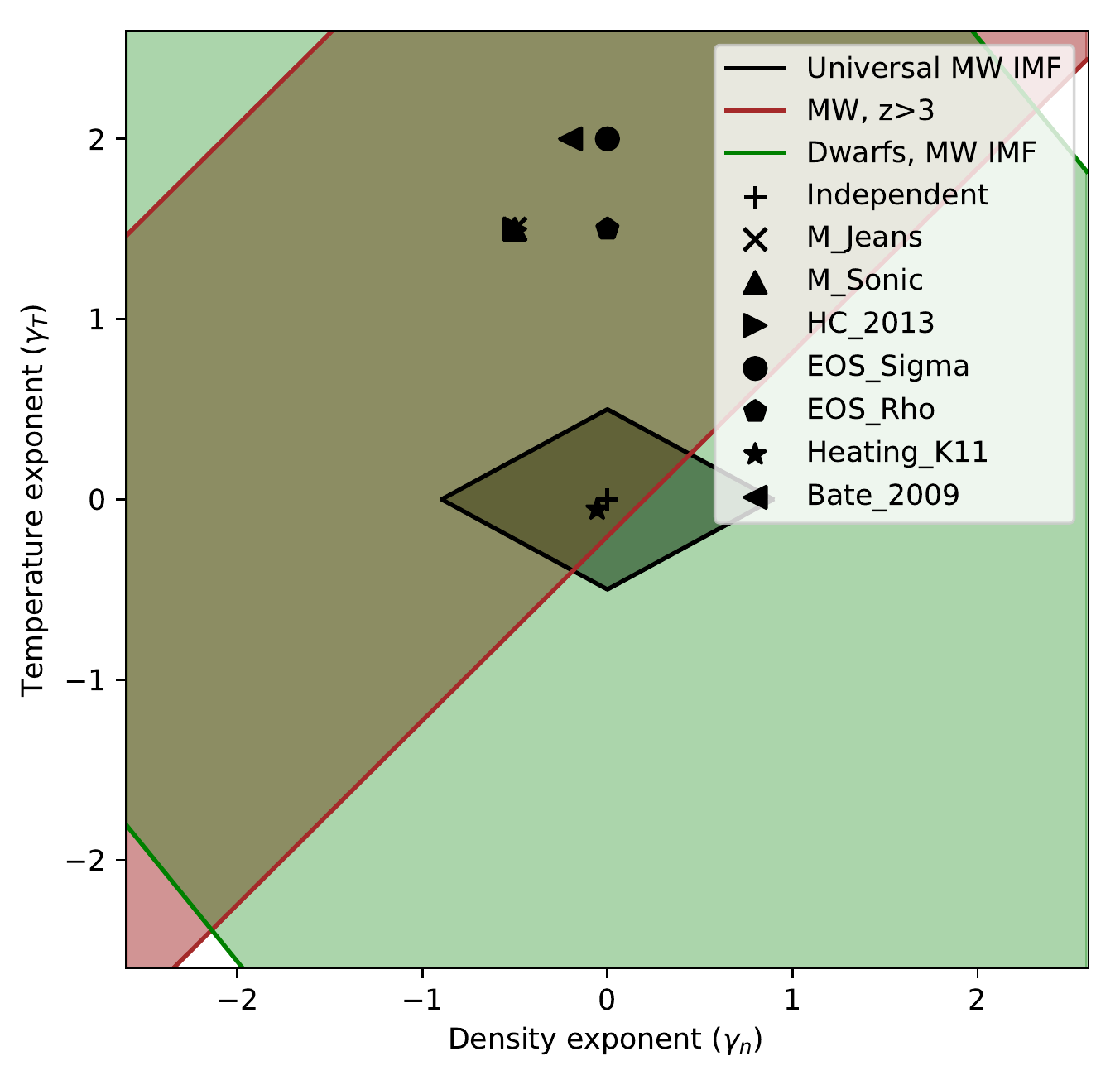}
\includegraphics[width=0.33\linewidth]{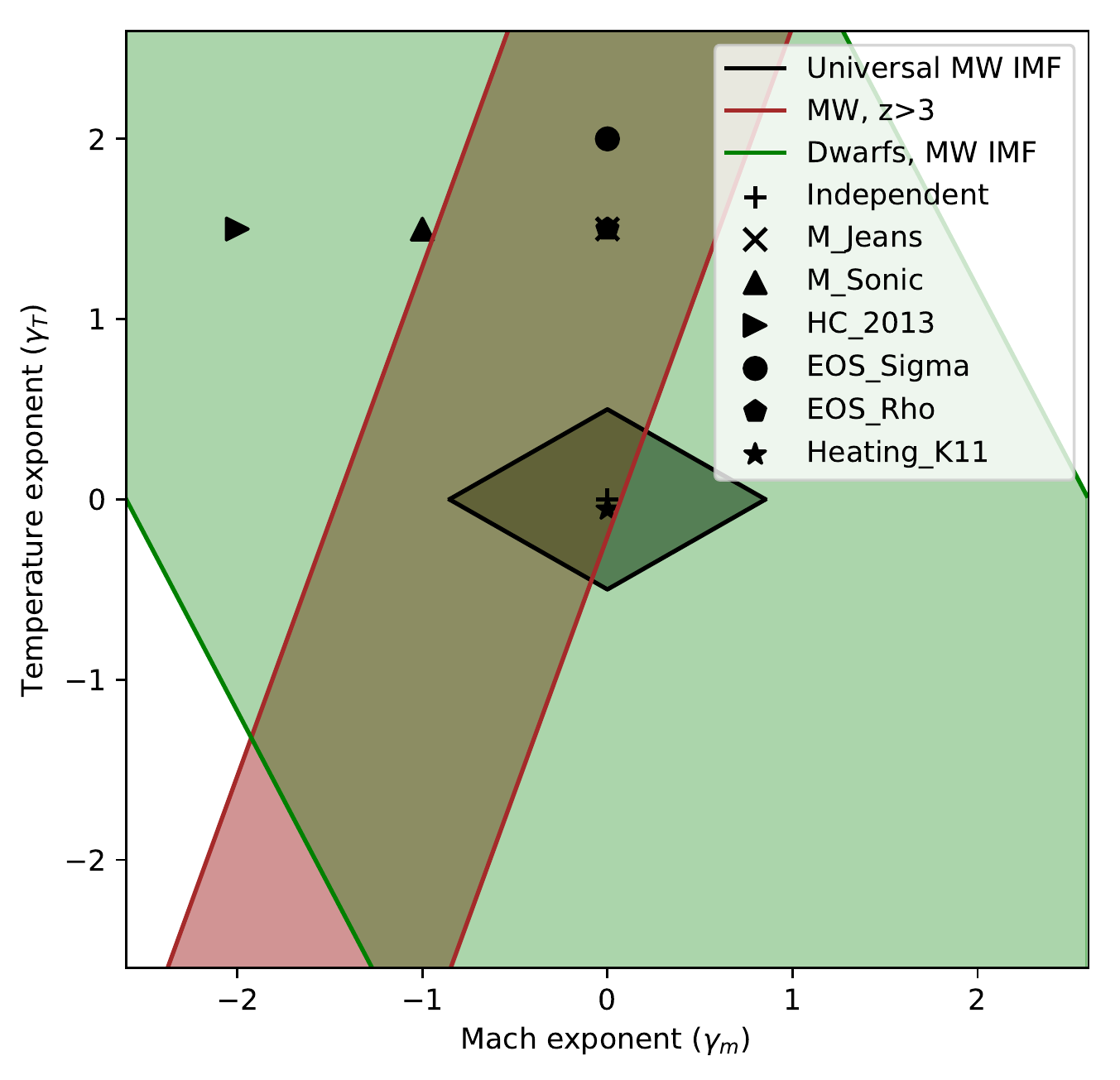}
\includegraphics[width=0.33\linewidth]{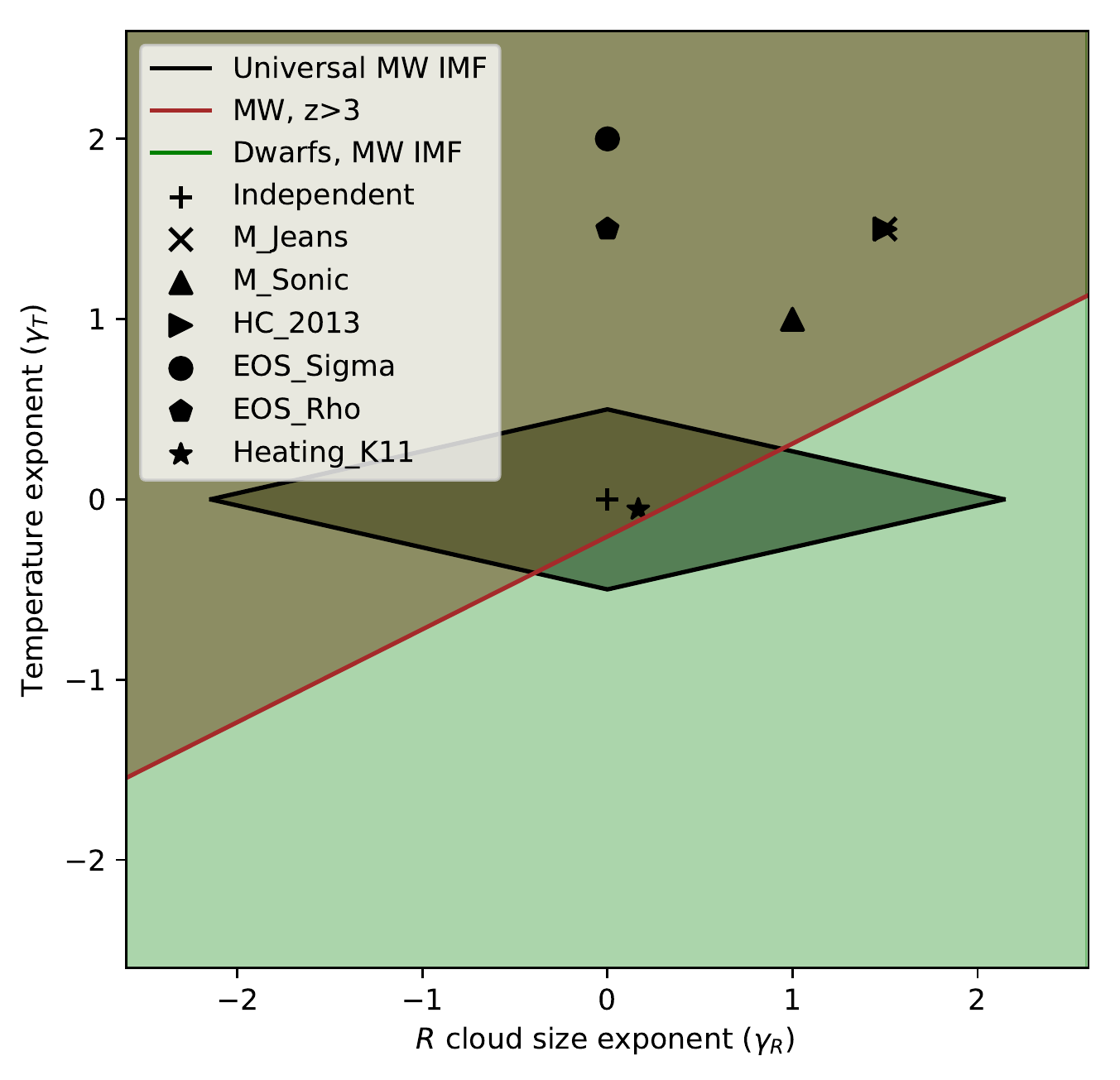}\\
\vspace{-0.15cm}
\includegraphics[width=0.33\linewidth]{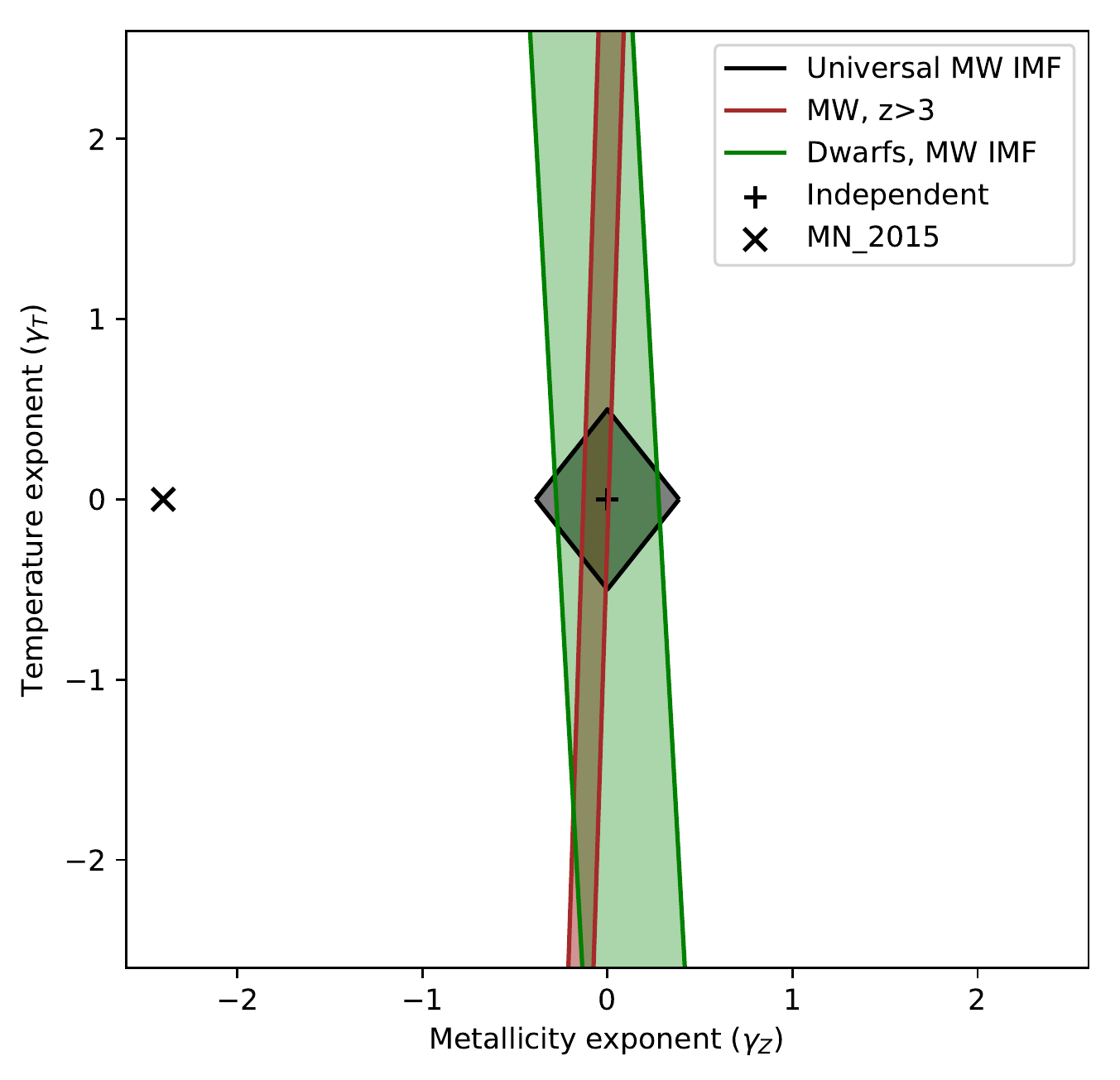}
\includegraphics[width=0.33\linewidth]{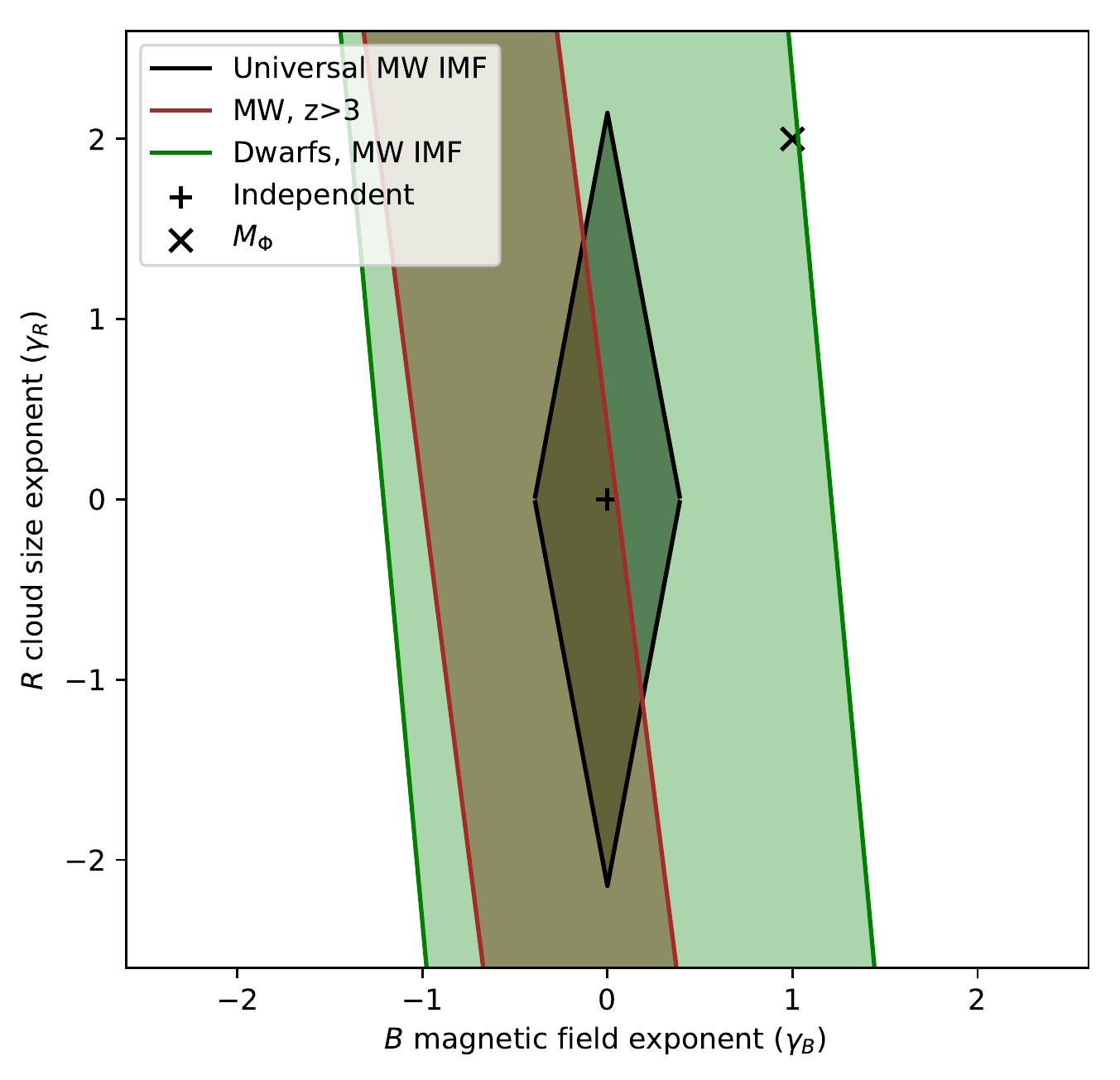}
\vspace{-0.4cm}
\caption{Power-law exponents for the density, temperature, metallicity, Mach-number, size and magnetic field in Equation \ref{eq:mu_powerlaw} that satisfy IMF universality in the MW and nearby dwarf galaxies (see Table \ref{tab:contraints} for details on constraints and Table \ref{tab:galaxies} for the simulated galaxies). The shaded regions show the exponents that satisfy the different constraints, while symbols represent models from the literature (Table \ref{tab:SF_models}). From these we can infer that there is a fairly limited volume in the model space of Equation \ref{eq:mu_powerlaw} that satisfies MW universality, the constraints are especially stringent on the $\gamma_{Z}$ metallicity exponent.}
\label{fig:exponents_universal}
\vspace{-0.5cm}
\end {center}
\end{figure*}

\begin{figure*}
\begin {center}
\includegraphics[width=0.33\linewidth]{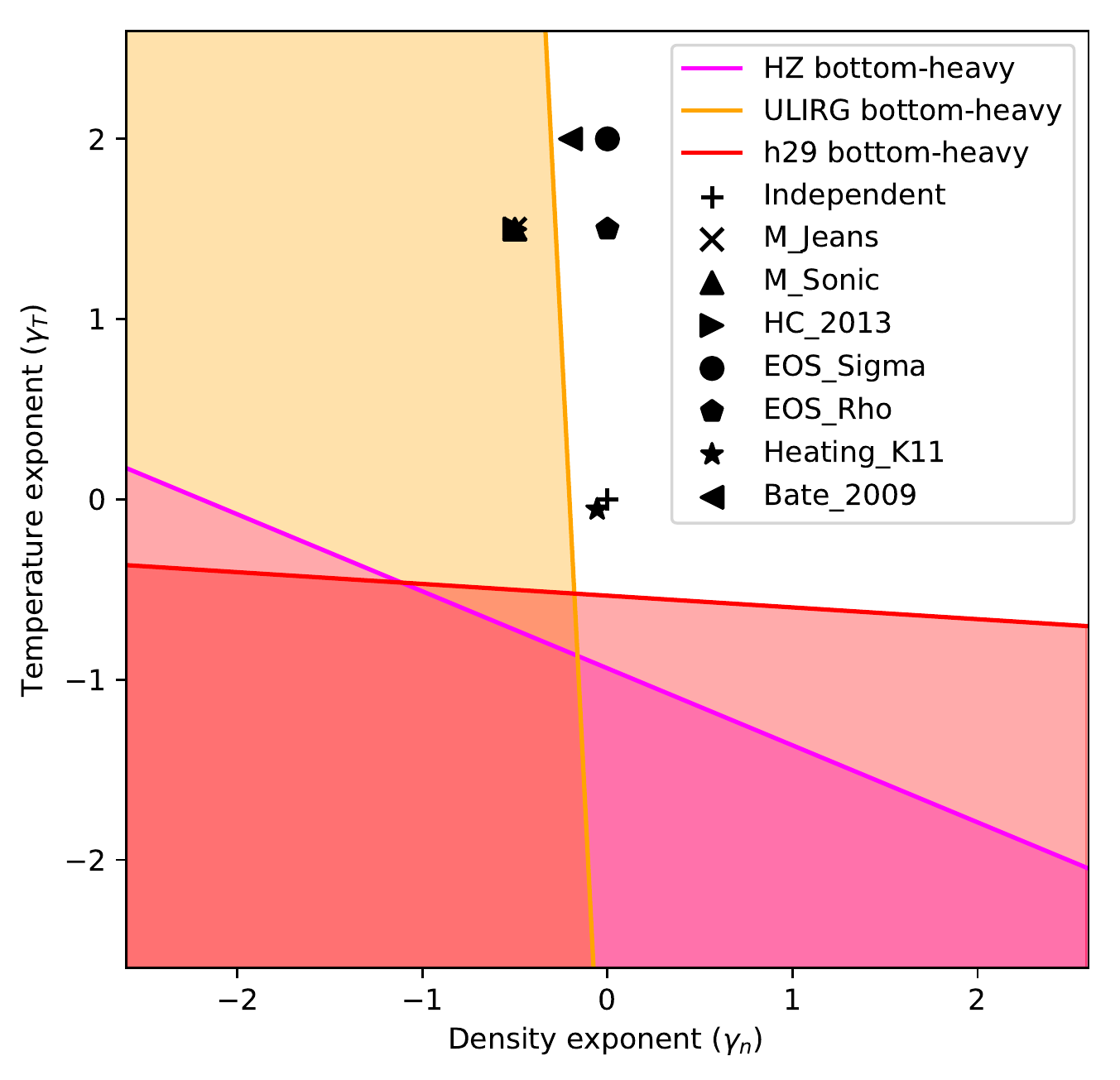}
\includegraphics[width=0.33\linewidth]{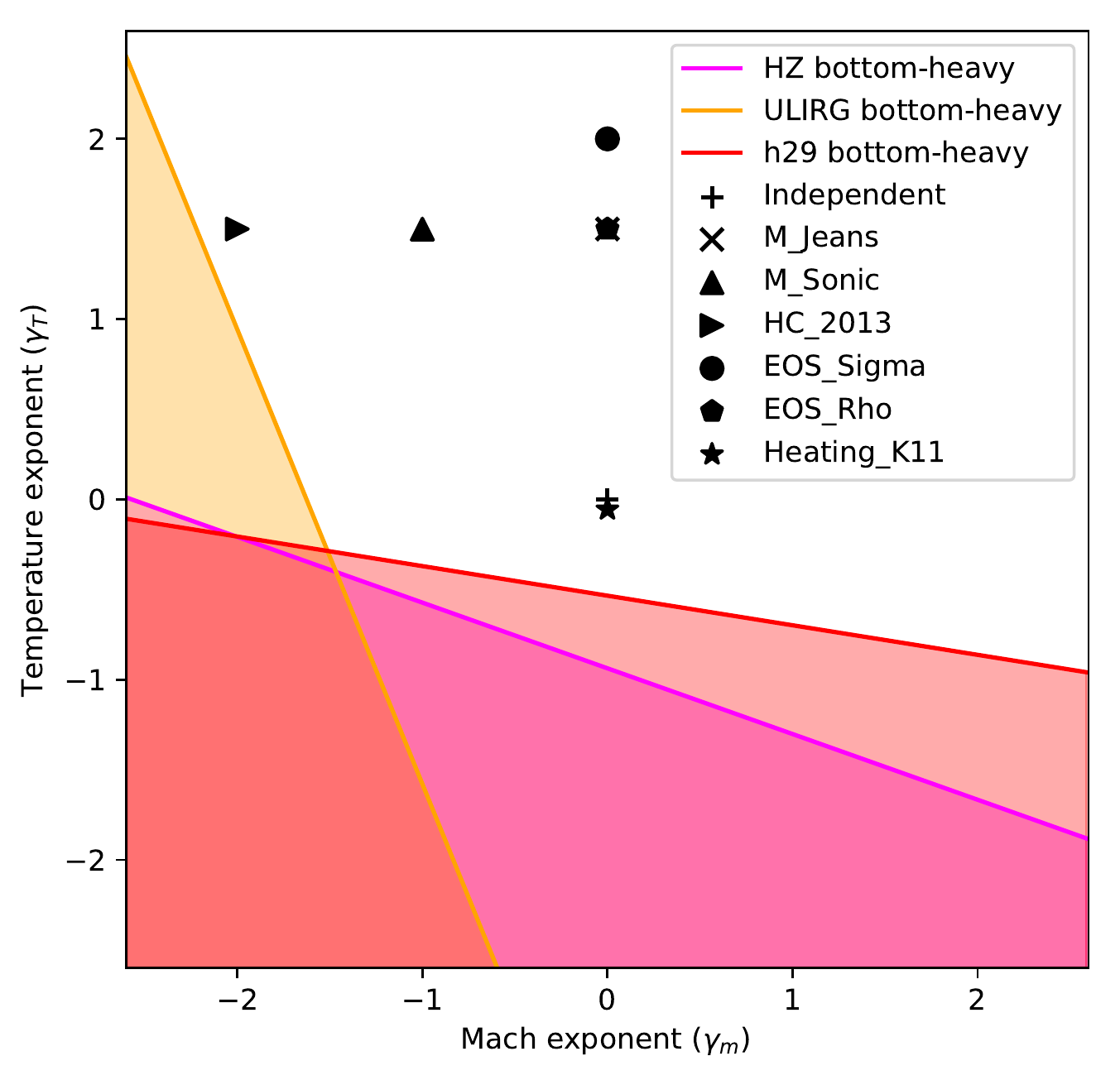}
\includegraphics[width=0.33\linewidth]{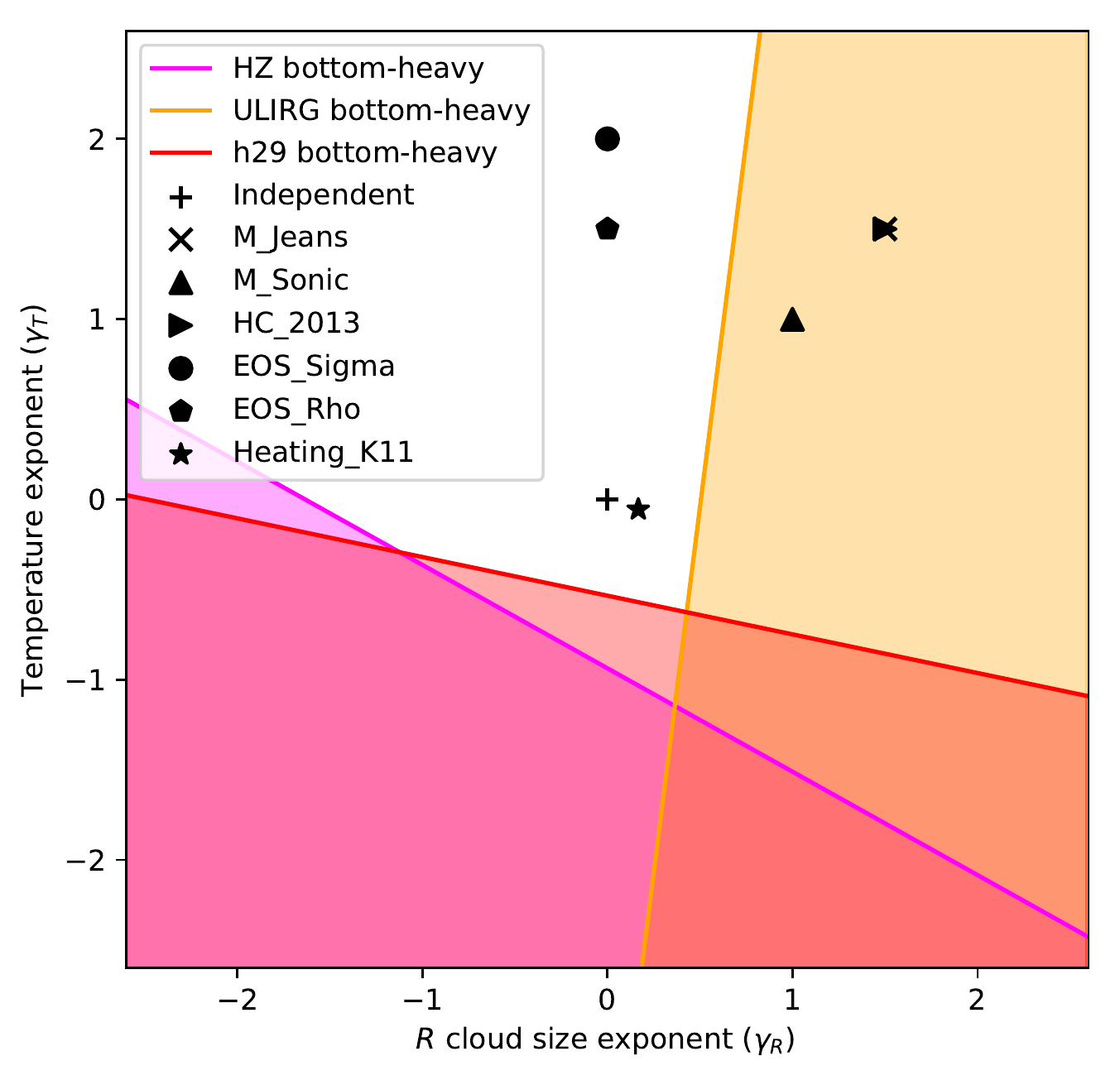}\\
\vspace{-0.15cm}
\includegraphics[width=0.33\linewidth]{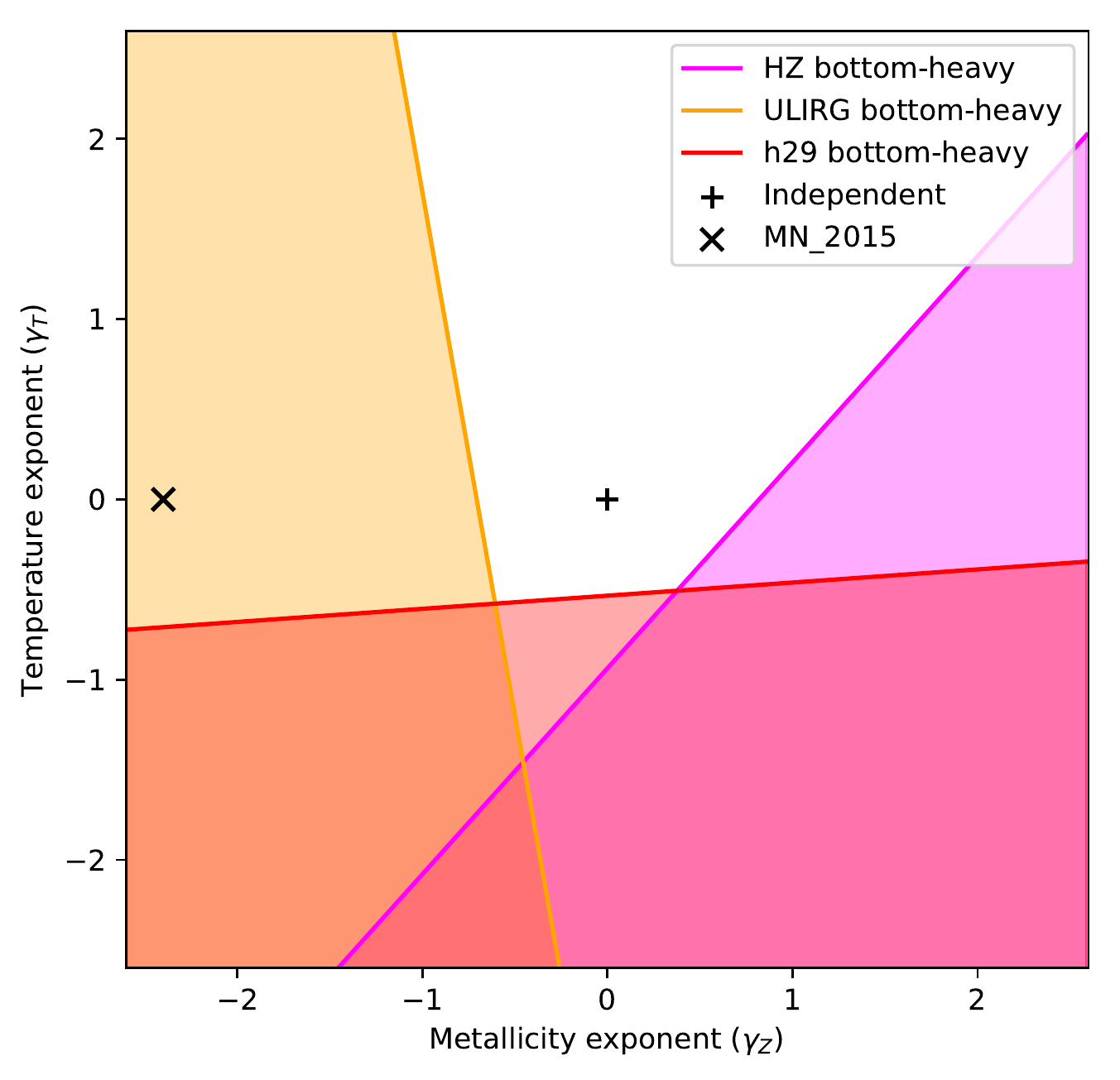}
\includegraphics[width=0.33\linewidth]{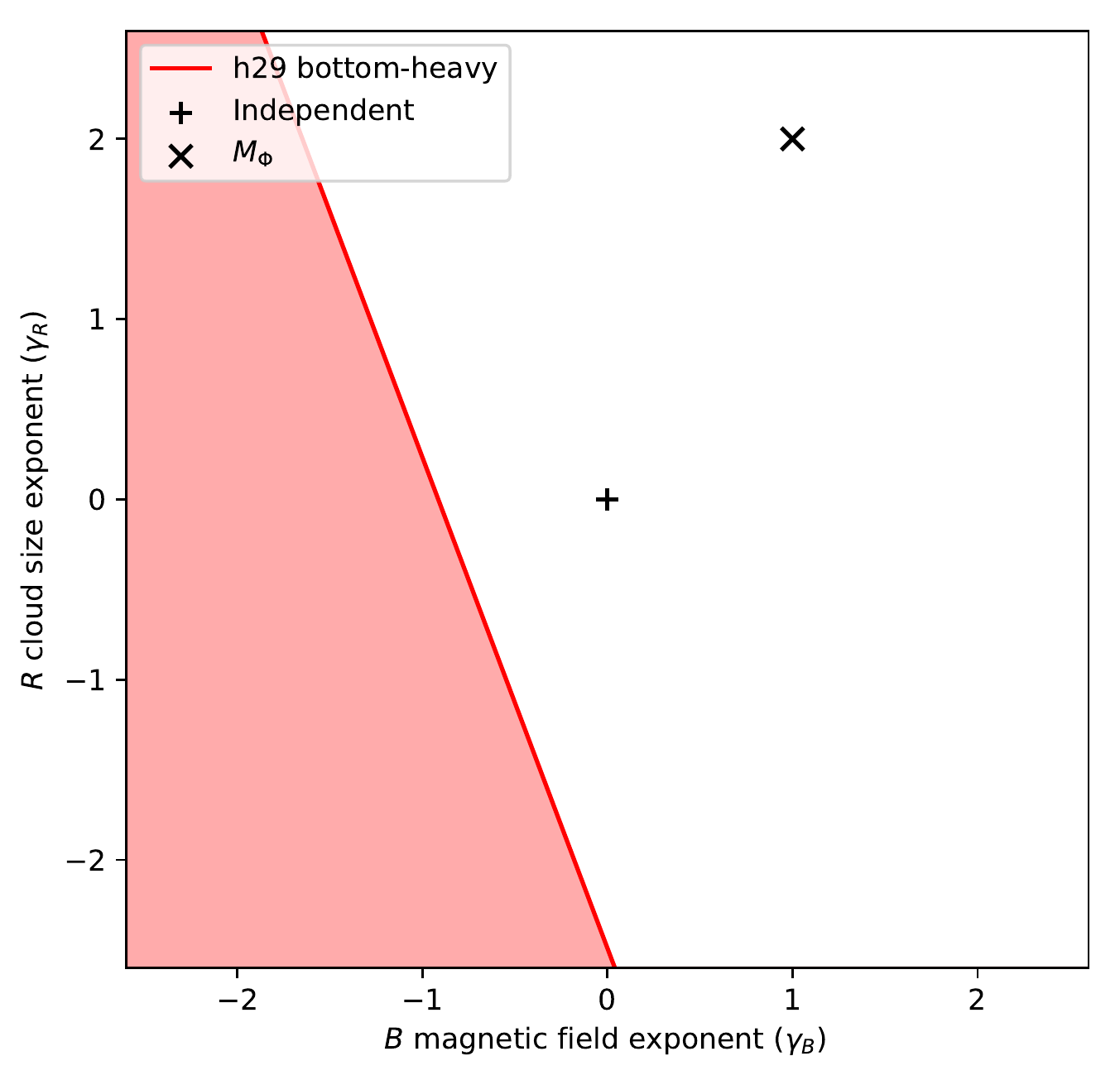}\\
\includegraphics[width=0.33\linewidth]{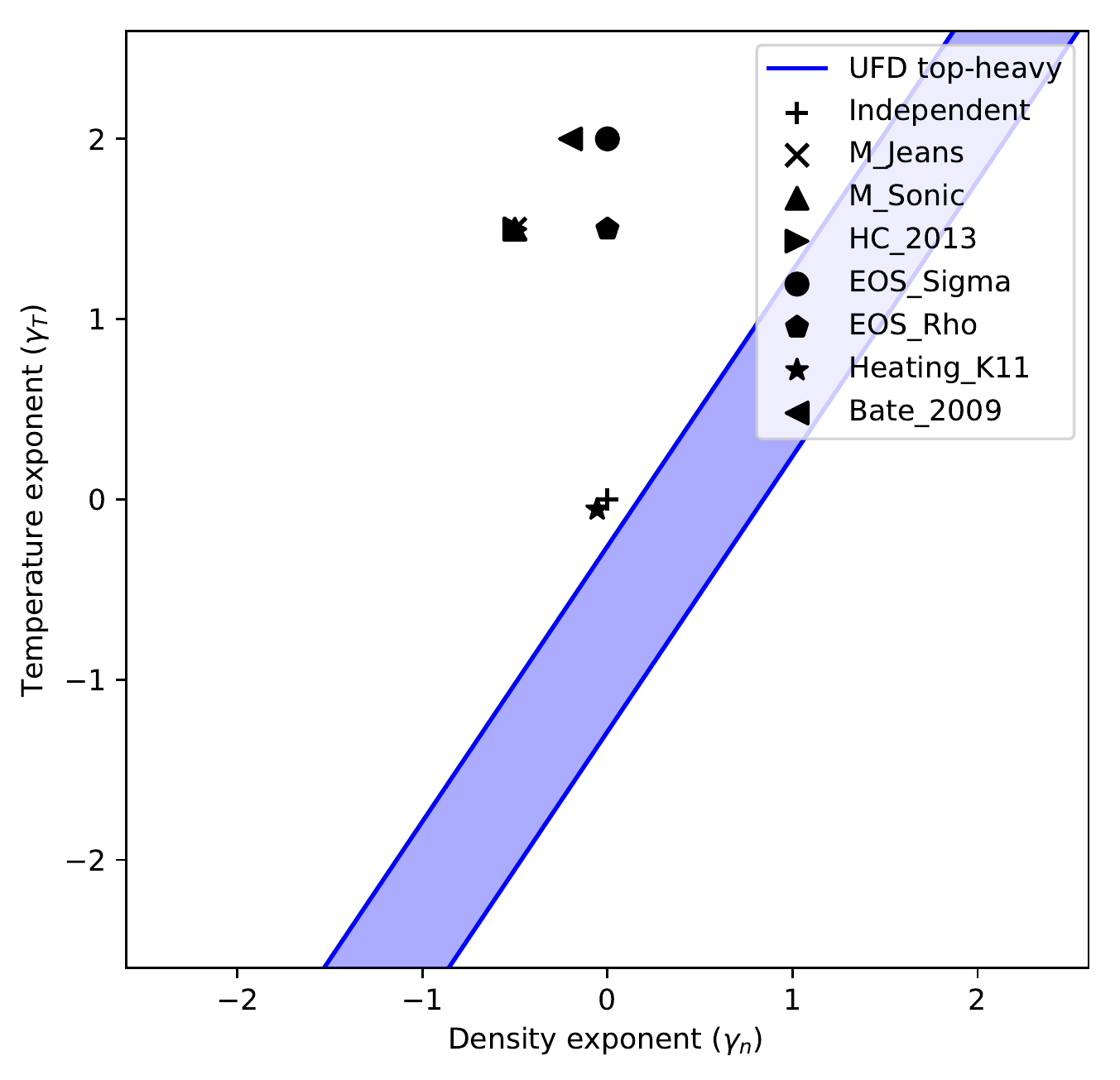}
\includegraphics[width=0.33\linewidth]{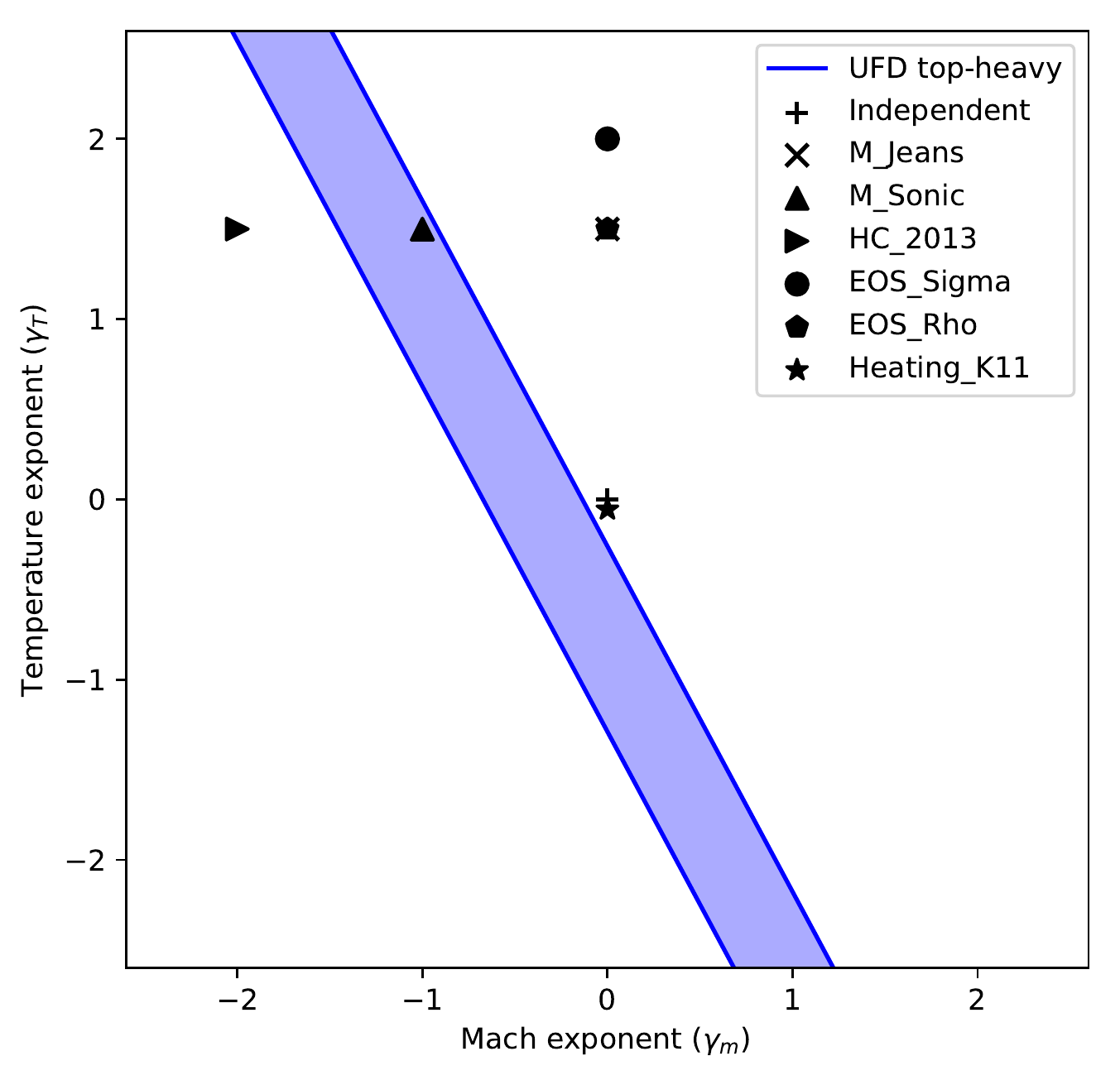}\\
\vspace{-0.15cm}
\includegraphics[width=0.33\linewidth]{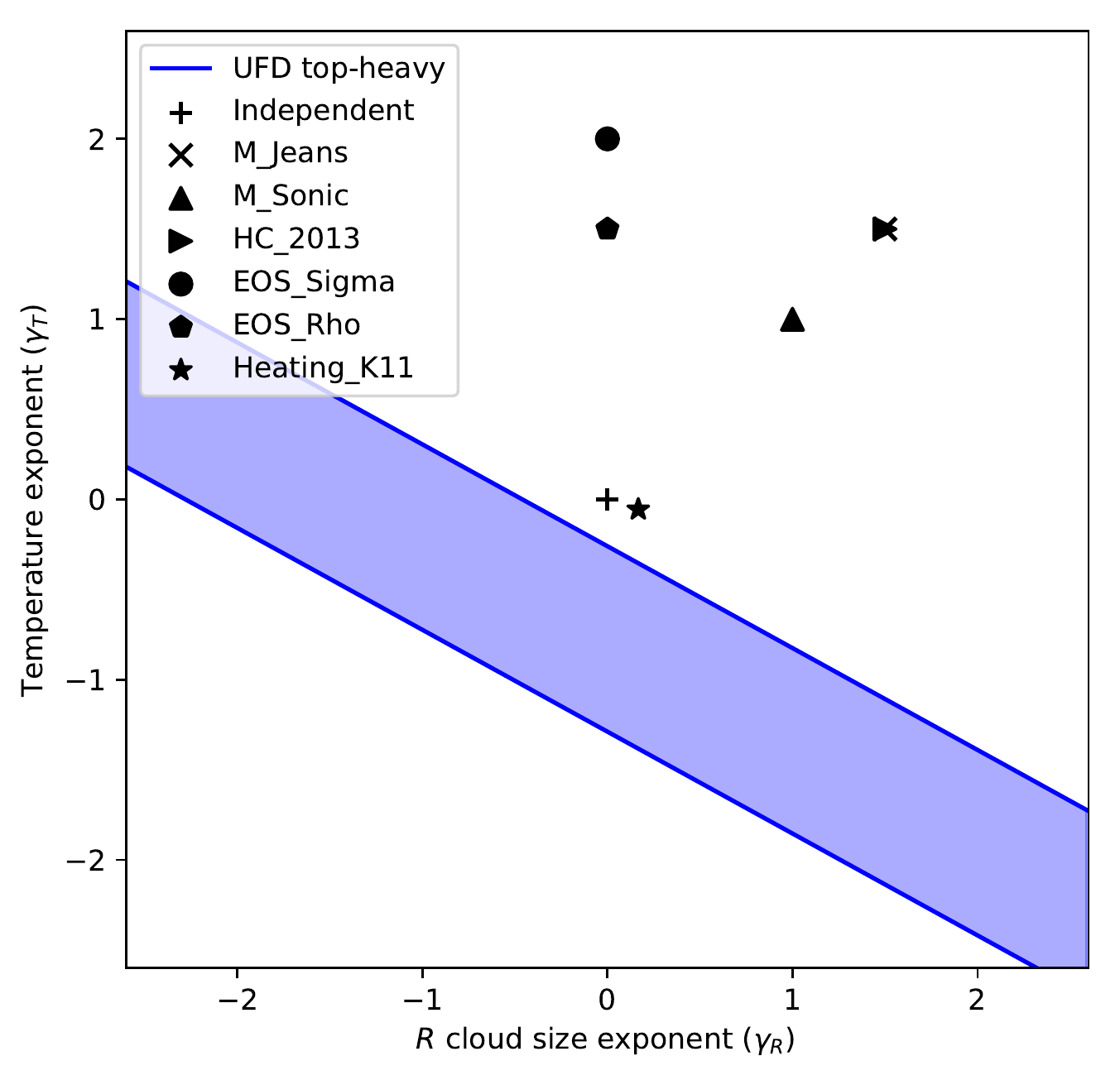}
\includegraphics[width=0.33\linewidth]{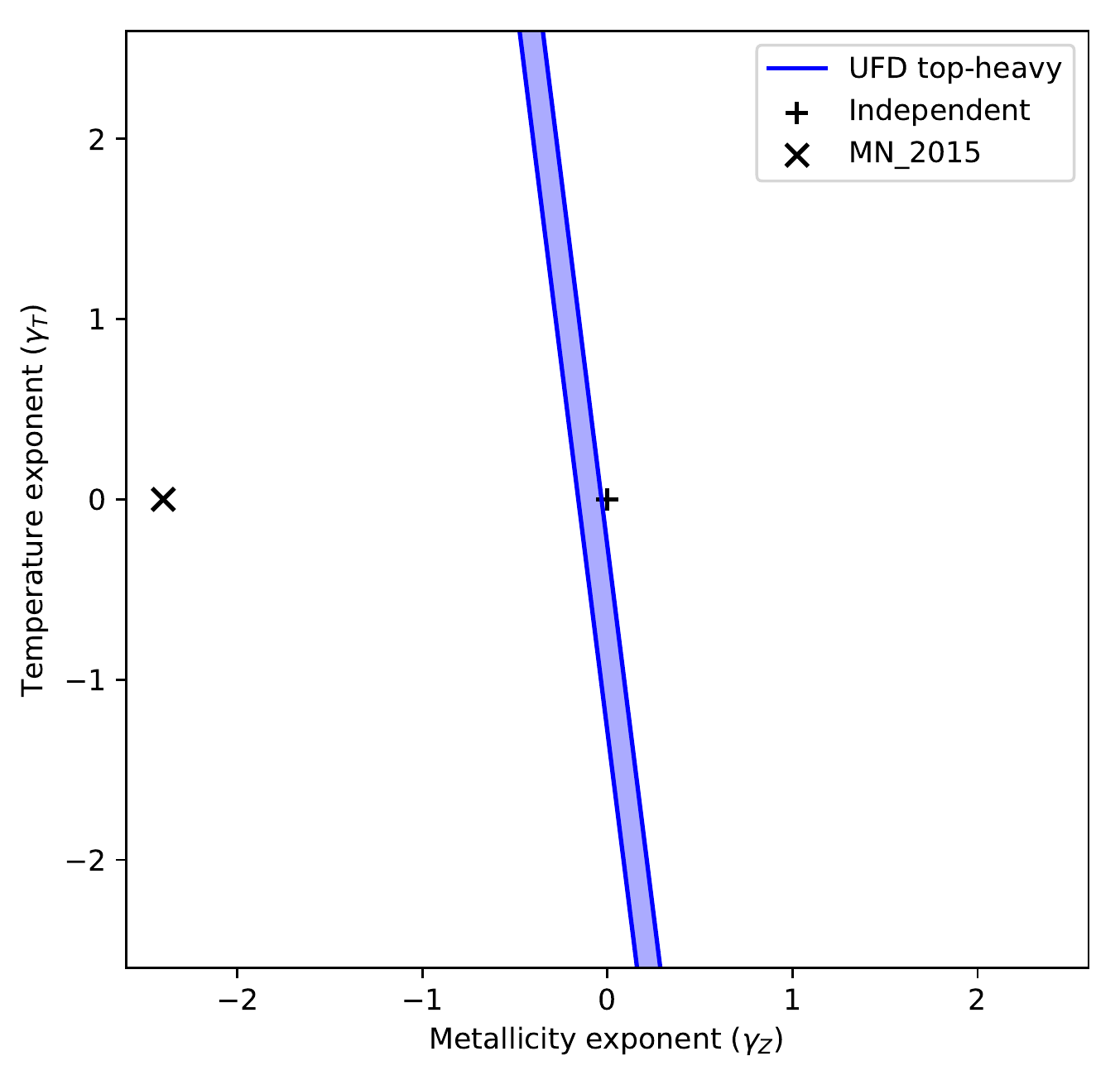}
\vspace{-0.4cm}
\caption{Power-law exponents for Equation \ref{eq:mu_powerlaw} that reproduce the inferred bottom-heavy IMF for early type galaxies (top 2 rows) and the inferred top-heavy IMF for ultra faint dwarf galaxies (bottom 2 rows), similar to Figure \ref{fig:exponents_universal}.Note that UFD proxy (\textbf{m10xf\_14706}) as well as one of our simulated early type galaxies (\textbf{z5m12c}) did not include magnetic fields, hence they provide no constraints on the $\gamma_B$ exponent. It is clear that the models in the literature fail to reproduce the bottom-heavy IMF for simulated galaxies, but some can satisfy the constraints when using canonical ULIRG values.}
\label{fig:exponents_vars}
\vspace{-0.5cm}
\end {center}
\end{figure*}

In this appendix we show how the individual constraints affect the available parameter space for the IMF models defined by Equation \ref{eq:mu_powerlaw}. Figure \ref{fig:exponents_universal} show that the requirement for low galactic scatter in the $\mu$ characteristic IMF mass drastically reduce the available exponents. Requiring that old stellar populations have similar or slightly more massive $\mu$ further restricts this space, especially in the case of the metallicity exponent $\gamma_Z$. Meanwhile, Figure \ref{fig:exponents_vars} show that although a large volume of parameter space would reproduce the inferred bottom-heavy IMF in early type galaxies, few models in the literature can do so and only in case of using canonical ULIRG values instead of simulated galaxies.

\end{document}